\documentclass[aps,prb,superscriptaddress,citeautoscript,twocolumn,notitlepage,longbibliography]{revtex4-2}

\usepackage{graphicx}
\usepackage{dcolumn}
\usepackage{bm}

\usepackage{graphicx}
\usepackage{amsmath}
\usepackage{amssymb}
\usepackage[normalem]{ulem}
\usepackage{braket}
\usepackage{color}
\usepackage{gensymb}
\usepackage[colorlinks=True,linkcolor=red,citecolor=blue,urlcolor=blue]{hyperref}

\usepackage{dsfont}
\usepackage{physics}
\usepackage{stmaryrd}
\usepackage{soul}

\usepackage{makecell}
\usepackage{dsfont}
\usepackage{mathtools,esint}
\newcommand{\bea}{\begin{eqnarray*}}
\newcommand{\eea}{\end{eqnarray*}}
\newcommand{\bne}{\begin{equation*}}
\newcommand{\ede}{\end{equation*}}

\newcommand{\bnen}{\begin{equation}}
\newcommand{\eden}{\end{equation}}
\newcommand{\bean}{\begin{eqnarray}}
\newcommand{\eean}{\end{eqnarray}}
\newcommand{\bsen}{\begin{subequations}}
\newcommand{\esen}{\end{subequations}}

\newcommand{\bna}{\begin{array}}
\newcommand{\eda}{\end{array}}
\newcommand{\bnm}{\begin{enumerate}}
\newcommand{\edm}{\end{enumerate}}
\newcommand{\bni}{\begin{itemize}}
\newcommand{\edi}{\end{itemize}}

\def\C{\mathbb C}

\def\R{\mathbb R}
\def\Z{\mathbb Z}

\def\im{{\rm Im}}

\renewcommand{\vec}[1]{\text{\boldmath{$ #1 $}}}

\DeclareMathAlphabet\mathbfcal{OMS}{cmsy}{b}{n}

\begin{document}

\title{Upper bound on the number of Weyl points born from a nongeneric degeneracy point}


\author{Gerg\H{o} Pint\'er}

\affiliation{Department of Theoretical Physics, Institute of Physics, Budapest University of Technology and Economics, M\H{u}egyetem rkp. 3., H-1111 Budapest, Hungary}

\author{Gy\"orgy Frank}

\affiliation{Department of Theoretical Physics, Institute of Physics, Budapest University of Technology and Economics, M\H{u}egyetem rkp. 3., H-1111 Budapest, Hungary}

\author{D\'aniel Varjas}

\affiliation{Department of Theoretical Physics, Institute of Physics, Budapest University of Technology and Economics, M\H{u}egyetem rkp. 3., H-1111 Budapest, Hungary}

\affiliation{Department of Physics, Stockholm University, AlbaNova University Center, 106 91 Stockholm, Sweden}

\affiliation{Max Planck Institute for the Physics of Complex Systems, N{\"o}thnitzer Strasse 38, 01187 Dresden, Germany}

\affiliation{Institute for Theoretical Solid State Physics, IFW Dresden and Würzburg-Dresden Cluster of Excellence ct.qmat, Helmholtzstr. 20, 01069 Dresden, Germany}

\author{Andr\'as P\'alyi}

\affiliation{Department of Theoretical Physics, Institute of Physics, Budapest University of Technology and Economics, M\H{u}egyetem rkp. 3., H-1111 Budapest, Hungary}

\affiliation{HUN-REN--BME Quantum Dynamics and Correlations Research Group, M\H{u}egyetem rkp. 3., H-1111 Budapest, Hungary}

\date{\today}

\begin{abstract}

Weyl points are generic and stable features in the energy spectrum of Hamiltonians that depend on a three-dimensional parameter space.
Non-generic isolated two-fold degeneracy points, such as multi-Weyl points, split into Weyl points upon a generic perturbation that removes the fine-tuning or protecting symmetry.
The number of the resulting Weyl points is at least $|Q|$, where $Q$ is the topological charge associated to the non-generic degeneracy point.
    Here, we show that such a non-generic degeneracy point also has a birth quota, i.e., a maximum number of Weyl points that can be born from it upon any perturbation. The birth quota is a local multiplicity associated to the non-generic degeneracy point, an invariant of map germs known from singularity theory. This holds not only for the case of a three-dimensional parameter space with a Hermitian Hamiltonian, but also for the case of a two-dimensional parameter space with a chiral-symmetric Hamiltonian. 
We illustrate the power of this result for band structures of two- and three-dimensional crystals.
Our work establishes a strong and powerful connection between singularity theory and topological band structures, and more broadly, parameter-dependent quantum systems. 
\end{abstract}

\maketitle

\section{Introduction}
Weyl semimetals are a class of topological materials whose electronic band structure exhibits pointlike linear band-touchings~\cite{Neumann,Herring,Armitage}.
These Weyl points are stable and generic, requiring no fine-tuning or symmetries.
Crystal symmetries, however, can stabilize isolated non-generic two-fold degeneracy points, such as multi-Weyl points, in the electronic band structure of three-dimensional solids \cite{ChenFang_multiweyl,TiantianZhang,Hirschmann,ZhiMingYu}. 
If the symmetry is broken, e.g., by changing an external magnetic (Zeeman) field or applying mechanical strain, the non-generic degeneracy point splits into multiple Weyl points~\cite{ChenFang_multiweyl}. 
As the symmetry-breaking perturbation is switched on gradually, these newly born Weyl points follow continuous trajectories in the Brillouin zone, originating from the original degeneracy point.

For a given non-generic degeneracy point, how many newborn Weyl points are allowed?
From topological charge conservation and the generic character of Weyl points, it follows that the minimum number of newborn Weyl points upon a generic perturbation is the absolute value $|Q|$ of the topological charge $Q$ associated to the non-generic degeneracy point (e.g. the Chern number in 3D).
The same consideration implies that the number of newborn Weyl points may also be $|Q|+2M$ with $M$ being a positive integer, in such a way that $M$ of the excess Weyl points have unit positive charge and $M$ have unit negative charge, hence the sum of the charges of the newborn Weyl points equals $Q$.
Is there also a `birth quota', i.e., an upper bound of the number of newborn Weyl points? 
This is a fundamental question, which has only been studied to a limited extent in the mathematical physics literature~\cite{Teramoto}, and the implications to realistic physical systems remain unexplored.

In this work, we answer this question positively, and show that 
such a non-generic degeneracy point does have a birth quota,
which turns out to be the so-called \emph{local multiplicity} known in singularity theory. 
Our result is not exclusive to Weyl points in three dimensions (3D), we also obtain an analogous result for two-dimensional (2D) crystals with chiral symmetry illustrated on the example of bilayer graphene, as well as a minimal example in 1D (Fig.~\ref{fig:z3}).
We emphasize that our analysis covers the effect of \emph{any} possible perturbation of the band structure. 
While this may sound surprising, it is possible to derive such general results using singularity theory.

We showcase the power of these results on quasiparticle (electronic, photonic, phononic) band structures: we compute the birth quota of all four types of two-fold degeneracies stabilised by the 230 crystalline space groups \cite{ZhiMingYu} (Table \ref{tab:multiplicities}).
However, the notion of the birth quota is more generic: it is applicable to quantum systems controlled by external parameters, such as interacting spin systems \cite{Wernsdorfer,Bruno,Scherubl,Frank,FrankDensity} or quantum circuits \cite{Fatemi,FrankTeleportation}; more generally, it is applicable to any physical system that is described by a matrix, e.g., linearly coupled mechanical oscillators \cite{Guba} or electromagnetic modes.

\section{Example: Bilayer graphene}
\begin{figure}
	\begin{center}
		\includegraphics[width=0.9\columnwidth]{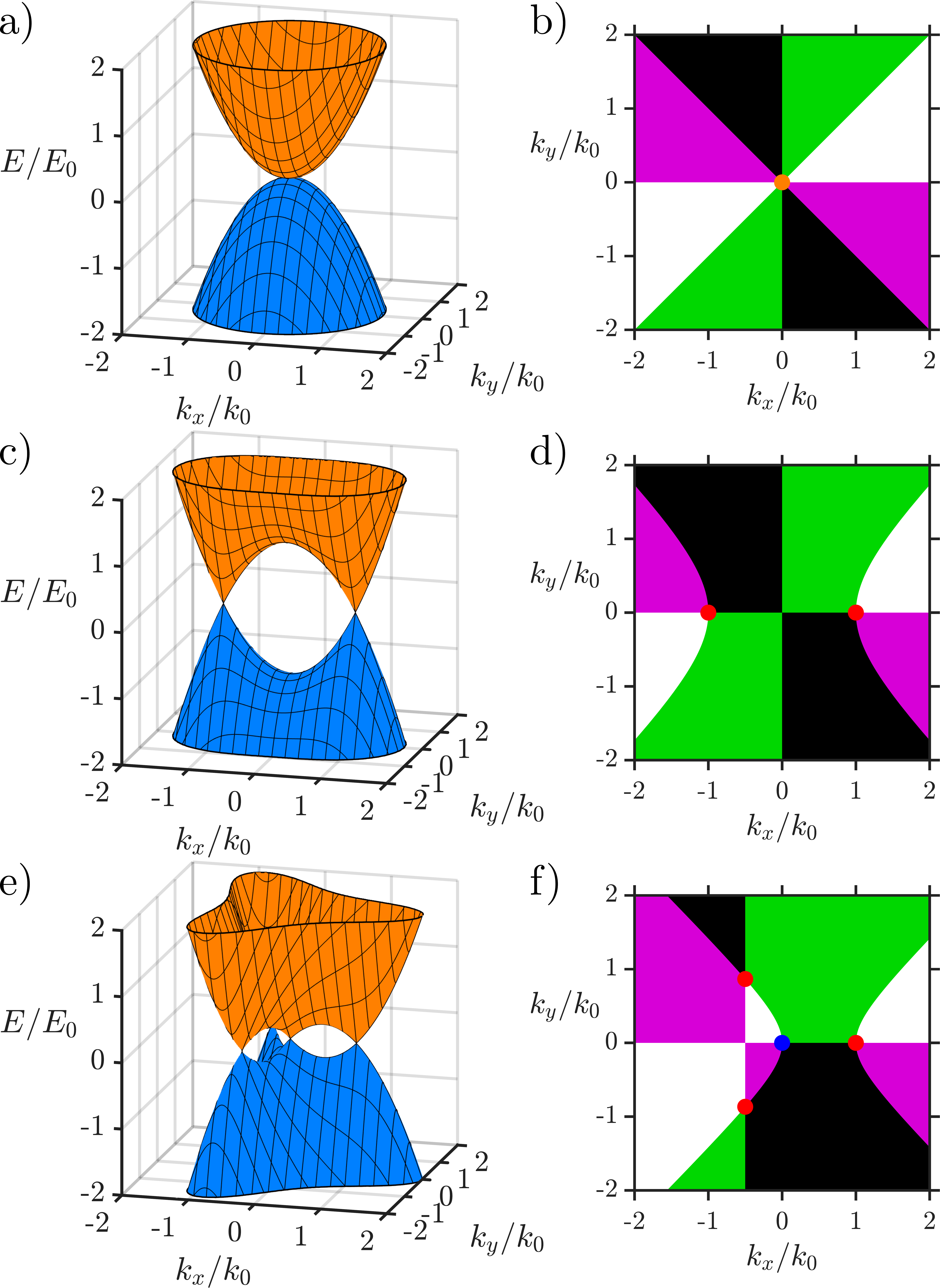}
	\end{center}
	\caption{Birth of 2D Weyl points from a non-generic degeneracy point in bilayer graphene.
	a-b) Non-generic degeneracy point. a) Dispersion relation of the valence band (blue surface) and conduction band (orange surface) based on Eq.~\eqref{eq:bilayer}. b) Colored plane shows the sign pattern of the effective Hamiltonian map $h$ of bilayer graphene, see Eq.~\eqref{eq:bilayermap}.
	White denotes $(+,+)$ regions, i.e., where both 
	components of $h(k_x,k_y)$ are positive. 
	Gray: $(-,+)$. Black: $(-,-)$. Green: $(+,-)$. Degeneracies are located where all four colors meet and their charge is equal to the signed number of  white-gray-black-green cycles as we go around them counter clockwise.
	c-d) Two 2D Weyl points born due to mechanical strain, both having charge $+1$, see Eq.~\eqref{eq:strain}. 
	e-f) Four 2D Weyl points born due to trigonal warping, see Eq.~\eqref{eq:trigonalwarping}. The three outer 2D Weyl points has charge $+1$, while the inner one has charge $-1$.
	The energy scale is $E_0=\hbar^2k_0^2/2m$, with the wave vector scale $k_0=\sqrt{2mw}/\hbar$ for c-d) and $k_0=2mv_3/\hbar$ for e-f). 
	The non-generic degeneracy point of a-b) is scale-invariant.
	\label{fig:w2}}
\end{figure}

We illustrate the birth of Weyl points from a non-generic degeneracy point by the well-known example of electrons in bilayer graphene (Fig.~\ref{fig:w2}).
This is a two-dimensional crystal, whose simplest band-structure models have chiral (a.k.a. sublattice) symmetry, which protects the accidental degeneracies (2D Weyl points) in the two-dimensional Brillouin zone.
A simple tight-binding model of delocalized electrons in graphene yields the following envelope-function Hamiltonian, valid in the vicinity of the high-symmetry point $K$ (and $K'$) of the Brillouin zone \cite{McCann,Peterfalvi}:
\bean
\label{eq:bilayer}
    H &=& \frac{\hbar^2}{2m}\begin{bmatrix}
    0 & (k_x-ik_y)^2 \\
    (k_x+ik_y)^2 & 0
    \end{bmatrix}.
\eean
Here, $k_{x,y}$ are momentum components measured from $K$, and the matrix represents a combined layer-sublattice degree of freedom.
The effective mass is $m\approx 0.035\ m_\text{e}$ \cite{McCann,Peterfalvi}.

This Hamiltonian exemplifies a non-generic degeneracy point in 2D, located at $(k_x,k_y) = 0$ with
dispersion $E_\pm =\pm \hbar^2k^2/2m$ quadratic in  momentum (Fig.~\ref{fig:w2}a).
To illustrate the structure of this degeneracy point, we first decompose the Hamiltonian with Pauli matrices, $H\propto \left(k_x^2-k_y^2\right) \sigma_x + 2k_xk_y\sigma_y$, where chiral symmetry forbids the $\sigma_z$ term~\cite{Asboth}.
It also forbids a $\sigma_0$ term, but as it does not affect the band-degeneracies, we ignore such terms in the following.
This allows us to reinterpret the Hamiltonian as the dimensionless
\emph{effective Hamiltonian map}
\begin{equation}
\label{eq:bilayermap}
h: \R^2 \to \R^2, (k_x,k_y) \mapsto (k_x^2-k_y^2, \ 2k_x k_y).
\end{equation} 
We plot how the signs (color coded) of the value of this map change on the $(k_x,k_y)$ plane in Fig.~\ref{fig:w2}b.
The meeting point of all colors defines the degeneracy point, and the color pattern in its neighborhood shows that this point has a winding number (or charge) of $Q_\text{2D} = 2$.

One type of perturbation to this degeneracy point is mechanical strain: applying it along $x$ adds the  perturbation term
\bean
    \label{eq:strain}
    H_\text{s} = -\begin{bmatrix}
    0 & w\\
    w & 0
    \end{bmatrix}
\eean
to $H$; strain of 1\% yields $w\approx$ 6 meV. \cite{Mucha}.
The effect of the strain is shown in Fig.~\ref{fig:w2}c,d: the non-generic degeneracy point is split to two 2D Weyl points at
\bean
\label{eq:strainpoints}
\vec k_{\pm}=\left(\pm\frac{\sqrt{2mw}}{\hbar },0\right),
\eean
both having charge $Q_\text{2D} = 1$, i.e., the total charge is conserved upon perturbation.

Another type of perturbation appears when the skew interlayer hopping is taken into account, causing 
\emph{trigonal warping}. 
This perturbation term is described by
\bean
\label{eq:trigonalwarping}
    H_\text{tw} = -\hbar v_3\begin{bmatrix}
    0 & k_x+ik_y\\
    k_x-ik_y & 0
    \end{bmatrix},
\eean
where $v_3\approx 10^5\ \frac{\text m}{\text s}$  \cite{McCann,Peterfalvi}. 
The effect of this perturbation is shown in Fig.~\ref{fig:w2}e,f: the non-generic degeneracy point of Fig.~\ref{fig:w2}a,b is split to 4 2D Weyl points:
3 having charge $Q_\text{2D} = 1$, located at
\bean
\label{eq:warpingpoints}
\vec k_n=\frac{2mv_3}{\hbar}\left(\cos\frac{2\pi n}{3},\sin\frac{2\pi n}{3}\right)\hspace{6mm}n\in\{1,2,3\},
\eean
and 1 having $Q_\text{2D} = -1$, located at $\vec k_4=(0,0)$. 
Again, the charge of the degeneracy point of $H$ is conserved by the perturbation. 

The observation that the non-generic degeneracy point at $K$ can split to 2 or 4 2D Weyl points triggers the question: is there a perturbation of $H$ such that the number of newborn 2D Weyl points is different from 2 or 4?
One implication of our analysis below is that the answer is `no', i.e., Fig.~\ref{fig:w2} covers all possibilities. 
The lower bound (2) is governed by topological charge conservation, whereas the upper bound (4), i.e., the `birth quota', follows from our analysis below.

Importantly, the topological charge and the birth quota can be defined here because chiral symmetry and the zero-energy nature of the degeneracy point together ensure that the source space and the image space of the effective Hamiltonian map $h: \R^2 \to \R^2$ has equal dimension (that is, 2). This must be true both with and without the perturbation. In case of monolayer or multilayer graphene, this condition is satisfied for zero-energy degeneracies if the tight-binding Hamiltonian has chiral symmetry, which is indeed the case in the bilayer example described above.
Note that in bilayer graphene, chiral symmetry can be broken by, e.g., an out-of-plane electric field \cite{McCann}, spin-orbit interaction \cite{KaneMele}, etc.

\section{Minimal 1D model}\label{s:minmod} 
\begin{figure}
	\begin{center}
		\includegraphics[width=0.9\columnwidth]{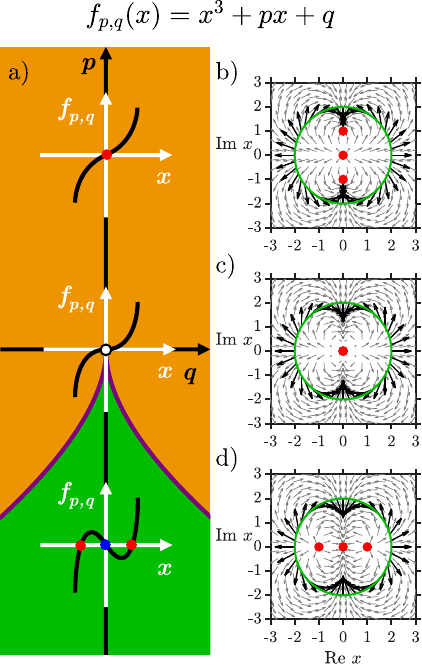}
	\end{center}
	\caption{
	Number of generic roots (`1D Weyl points') born from the non-generic root of $f(x) = x^3$ is lower-bounded by the charge of the root (which is 1) and upper-bounded by the local multiplicity of the root ($\text{mult}_0 f = 3$).
	a) Phase diagram of perturbations of $f(x) = x^3$
	of the form $f_{p,q}(x) = x^3+p x + q$.
	Orange: perturbation has one generic root.
	Green: perturbation has three generic roots.
	Purple line: perturbation has one generic root and one non-generic root. 
	b) Complexified perturbation $f_{\C,p,q}$ with $p=1$, $q=0$.
	c) Complexified map $f_\C(x) = x^3$, exhibiting a single root with a winding number or local multiplicity $\text{mult}_0 f = 3$.
	d) Complexified perturbation $f_{\C,p,q}$ with $p=-1$, $q=0$.
	The complex values of the function are represented as arrows, so phase windings can be read off.
	b), d) exhibits three complex roots with winding number $1$ each, even though the real function $f_{p,q}$ has b) one real root and d) three real roots. 
	\label{fig:z3}}
\end{figure}

First, we consider a minimal mathematical model of the birth quota effect: the birth of generic roots (`1D Weyl points') of the polynomial $f:\R \to \R, \, f(x) = x^3$, from its non-generic root at the origin $x=0$.
Any generic perturbation of the form $f_t(x)=x^3+t_1x^2+t_2x +t_3$ has one or three real roots, and all of them converge to $0$ as the vector of control parameters $t=(t_1, t_2, t_3)$ tends to 0. 
We draw such perturbations in Fig.~\ref{fig:z3}a, where, for simplicity, we set $(t_1,t_2,t_3) = (0,p,q)$.
In the case of three roots (green region), two of them have positive signs (i.e. the function is increasing, charge $+1$, red dot), and one has negative sign (the function is decreasing, charge $-1$, blue dot). 
In the case of one root (orange region), it has positive sign (charge $+1$).
Hence the sum of the charges is $+1$, same as the charge of the root of the unperturbed function $f$, exemplifying charge conservation. 

The number of the real roots changes when the control vector $t$ steps through the zero locus $\mathcal{D}=\{(t_1, t_2, t_3) \in \R^3 \ | \ D(t_1, t_2, t_3)=0 \}$ of the discriminant $D$ of $f$ (see Appendix~\ref{app:alg}). This \emph{discriminant set} $\mathcal{D}$ in the control space is illustrated as the purple solid line in Fig.~\ref{fig:z3}a.

Importantly, when we consider the complexification 
\footnote{For a real analytic function $f: \R^m \to \R^n$, its complexification is defined as the function $f_\C: \C^m \to \C^n$ that has the same power series as $f$.} 
$f_{\C,t} : \C \to \C$, the \emph{complex discriminant set} $\mathcal{D}_\C$ does not separate different root configurations: one can always find a path between different configurations such that the roots avoid collision in the complex plane. 
In fact, $f_{\C,t}$ has three distinct complex roots for any $t \in \C^3 \setminus \mathcal{D}_\C$.
This is illustrated by Fig.~\ref{fig:z3}b,c,d, displaying the (c) unperturbed map $f_{\C}$ and (b,d) two perturbed maps, and the roots of those $\C \to \C$ maps as red points.
The number of complex roots born upon perturbation from the root of $f_{\C}(x) = x^3$ is called the \emph{local multiplicity} $\text{mult}_0 f_{\C} = 3$ of $f_{\C}$ at the root $x=0$.
As illustrated in Fig.~\ref{fig:z3}c, this local multiplicity is the winding number of $f_{\C}$ evaluated on a small circle (green) around the root; 
the value 3 can be read off.

This winding number is also called the \emph{local degree} $\text{deg}_0 f_{\C}$, which is defined as the local degree $\text{deg}_0 f_{\C,\R}$ of the real-imaginary decomposition (`realification')
\bean
    f_{\C,\R}&:& \R^2 \to \R^2, \nonumber \\
    (u,v) &\mapsto& (\text{Re}(f_{\C}(u+i v)), \text{Im}(f_{\C}(u+iv))
\eean
of the complexification $f_{\C}$. Here, the explicit form of $f_{\C,\R}$ reads $f_{\C,\R}(u,v)=(u^3-3u v^2,3u^2v-v^3)$.
Since every generic root of $f_t$ is also a generic root of $f_{\C,t}$, we conclude that the number of roots born from the original root of $f$ at $x=0$ upon perturbation is upper-bounded by $\text{deg}_0 f_{\C,\R}$.

Note that higher-degree perturbations of $f$ can have more than three roots, but these extra roots are `born at infinity', not from the original root at $x=0$. 
For example, the root $x=1/t$ of the perturbed map \mbox{$f_t(x)=x^3-tx^4=-tx^3(x-1/t)$} goes to infinity as $t$ tends to 0.

In the rest of this paper, we outline the generalization of this minimal model, and use it to derive the birth quota of non-generic isolated two-fold degeneracy points.

\section{Effective Hamiltonian at a twofold degeneracy point} 
The models we study, e.g., tight-binding models of the electronic band structure, are described by a Hamiltonian map $H:M^m \to \mbox{Herm}(N)$.
Here, $M^m$ is an $m$-manifold, and $\mbox{Herm}(N)$ is the real vector space of $N \times N$ Hermitian matrices. 
We focus on the example when $M^m = \text{BZ}$ is the Brillouin zone of a 3D crystal ($m=3$), but will also comment on the 2D case ($m=2$), relevant for bilayer graphene described above.

A key quantity in our analysis of the birth quota, analogous to the function $f$ in the minimal model above, is the \emph{effective Hamiltonian map} $h$ associated to the non-generic isolated two-fold degeneracy point $P \in \text{BZ}$.
Denote the ordinal number of the degenerate levels by $i$ and $i+1$.
Then, for the eigenvalues of $H(P)$, it holds that $E_{i-1} < E_i= E_{i+1} < E_{i+2}$,  and $E_i < E_{i+1}$ holds in a neighborhood $U_P$ of $P \in \text{BZ}$. 
The effective Hamiltonian map $h$ is obtained by an exact Schrieffer--Wolff transformation \cite{Bravyi} at $P$, which provides a map from the neighborhood $U_P$ of the degeneracy point $P$ into the space of traceless Hermitian $2\times 2$ matrices. 
The latter matrix space is identified with $\R^3$ via the standard Pauli-matrix decomposition, $X \sigma_x+Y \sigma_y+Z \sigma_z \equiv (X, Y, Z)$, hence the Schrieffer--Wolff transformation yields the effective Hamiltonian map $h:U_P \to \R^3$.

The fact that $P$ is an isolated degeneracy point can be reformulated as $h^{-1}(0)=\{P\}$, i.e. the pre-image set of the origin $0 \in \R^3$ contains only one point. 
We assume that in the natural wave-vector coordinates of the BZ, measured from $P$ as the reference point, $h$ is an analytic map, that is, its Taylor series is convergent  and it produces $h$ \footnote{Note that our analysis also works in $\mathcal{C}^{\infty}$ (i.e. smooth) category, using the notion of \emph{map germs}, however the analytic condition simplifies the discussion.}.
This results in an $h: (\R^3,0) \to (\R^3,0)$ map, consisting of 3 locally convergent power series $h_1$, $h_2$ and $h_3$ of 3 variables, fulfilling $h(0) = 0$. 
Our following analysis works for $m=3$ in general, and also for $m=2$ assuming chiral symmetry, when the degeneracy is at zero energy. The latter conditions imply $h_3=0$ and provide a map $h: (\R^2,0) \to (\R^2, 0)$
\footnote{Similarly, spinless $P\mathcal{T}$ symmetry restricts the Hamiltonian to be real valued, yielding $h_2=0$. Hamiltonians with both chiral and $P\mathcal{T}$ symmetry correspond to the case $m=1$ with effective Hamiltonian map $h: (\R,0) \to (\R, 0)$.
Such symmetric 1D systems can provide a physical realization of the minimal example with $H(k) = f(k) \sigma_x$.}.

\section{Birth quota}
Up to now, we converted band-structure features to an $h:\R^3 \to \R^3$ analytic map.
This enables us to establish the birth quota of a non-generic degeneracy point using concepts and relations from singularity theory.
The \emph{local degree} $\deg_0 h$ of the effective Hamiltonian map $h: (\R^m, 0) \to (\R^m, 0)$ at $0$ is the \emph{global degree} of the normalised map (`pseudospin texture' \cite{TiantianZhang}) $\tilde{h}=\frac{h}{|h|}: S^{m-1}_{\epsilon} \to S^{m-1}$  defined on a sufficiently small sphere $S^{m-1}_{\epsilon}$ around the origin, see
\cite{milnortop, arnold, Eis, EisLev, mond-ballesteros} for details.
For $m=2$ the local degree is a winding number, and for $m=3$ it agrees with the first Chern number of the eigenstate corresponding to the $i$-th eigenvalue, see \cite{Asboth}.
For both cases, the local degree is often referred to as the (topological) charge of the degeneracy point.
We will also refer to the complexification $h_{\C}: (\C^m, 0) \to (\C^m, 0)$ of $h$, and its local degree $\deg_0 h_{\C}$.

The effect of a physical perturbation on the crystal electrons, e.g., mechanical strain or a change of the magnetic (Zeeman) field, is described in our framework as a \emph{deformation} of the effective Hamiltonian map $h$.
An unfolding of $h$ with $k$ \emph{control parameters} is an analytic map $\mathcal{H}: \R^3 \times \R^k \to \R^3 \times \R^k$ of the special form $\mathcal{H}(x, y, z, t)=(h_t(x, y, z), t) $, where $t \in \R^k$,  such that $h_0=h$. 
For fixed control parameters $t$, we call $h_t$ an analytic deformation, or simply deformation, of $h$.
To derive the birth quota of $h$, we will use the complex generalizations of these concepts:
a holomorphic unfolding $\mathcal{H}_{\C}$ and the corresponding complex deformation $h_{\C,t}$ of the complexification $h_{\C, 0}=h_{\C}$ are defined similarly to the real case above \cite{mond-ballesteros, arnold}.

We call the points of  $h_t^{-1}(0) \subset \R^3$ \emph{degeneracy points} of $h_t$, and the points of $h_{\C, t}^{-1}(0) \subset \C^3$ \emph{complex degeneracy points}. 
A complex degeneracy point $p$ is generic, and we call it a \emph{complex Weyl point}, if the Jacobian of $h_{\C, t}$ at $p$ has maximal rank, i.e., rank 3.
A real degeneracy point $p$ is generic, and we call it a (real) Weyl point, if the Jacobian of $h_t$ at $p$ has maximal rank. 
This happens if and only if $p$ is generic as a complex degeneracy point. 
At a real Weyl point $p$, the local degree is $\deg_p h_t=\pm 1$, determined by the sign of the Jacobian determinant, cf. \cite{ChenFang_multiweyl}.
At a complex Weyl point $p$, the local degree of $h_{\C, t}$ is always $\deg_p h_{\C, t}=1$.

Our goal is to characterize the birth of (real) Weyl points from a non-generic degeneracy point. 
Hence, we need to distinguish between Weyl points born from the original degeneracy point, and all other Weyl points.
This we already noted in the minimal model above, by pointing out that 1D Weyl points can be born at infinity.
To make this distinction here, we consider the complex case first. 
We take a spherical boundary $S^5_\epsilon \subset \C^3$ in the configuration space, centered at the origin, which we call the \emph{separator}, such that the only degeneracy point of $h_{\C}$ inside this sphere is the origin.
The separator bounds the closed ball $B^6_\epsilon \subset \C^3$.
When a deformation is applied continuously, the degeneracy points $h^{-1}_{\C,t}(0)$ follow continuous trajectories in the complex configuration space. 
Therefore, there is a neighborhood $\mathcal{U}_{\C}$ of the origin of the control space such that for all $t \in \mathcal{U}_{\C}$, (i) the degeneracy points born from the original degeneracy point do not reach the separator, and (ii) the degeneracy points of the undeformed map $h_{\C}$ that are outside of the separator do stay outside.
For convenience, we take a neighborhood $\mathcal{U}_{\C}$ that is an open ball centered at the origin of the control space.
Furthermore, from now on, we restrict the unfolding $\mathcal{H}_{\C}$ onto $B^6_\epsilon \times \mathcal{U}_{\C}$.

We define the \emph{complex discriminant set} $\mathcal{D}_{\C} \subset \mathcal{U}_{\C}$ as those control vectors $t$ for which $h_{\C,t}$ has non-generic complex degeneracy points in $B^6_\epsilon$.
That is, for any control vector $t \in \mathcal{U}_{\C} \setminus \mathcal{D}_{\C}$, all degeneracy points of $h_{\C,t}$ are complex Weyl points. 
A key observation is that the number of complex Weyl points within $B^6_\epsilon$, denoted as $\sharp h_{\C,t}^{-1}(0)$, is the same for any control vector $t \in \mathcal{U}_{\C} \setminus \mathcal{D}_{\C}$.
This is related to the fact that the complex codimension of the discriminant set $\mathcal{D}_{\C}$ is at least 1, and therefore the real codimension is at least 2.
This implies that $\mathcal{U}_{\C} \setminus \mathcal{D}_{\C}$ is path-connected, and hence the number $\sharp h_{\C,t}^{-1}(0)$ of complex Weyl points cannot change along any control trajectory in $\mathcal{U}_{\C} \setminus \mathcal{D}_{\C}$.
For details, we refer to Appendix~\ref{app:mult1} and \cite{mond-ballesteros}.
In conclusion, the number of preimages of $0$, $\sharp h_{\C,t}^{-1}(0)$ is a property of $h_{\C}$, and hence a property of $h$.
It is called the \emph{local multiplicity} of $h_{\C}$ ($\text{mult}_0 h_{\C}$), and also called the \emph{local multiplicity} of $h$ ($\text{mult}_0 h$).

In the band-structure context, the effective Hamiltonian map $h$ is a real map. 
We denote the set of real control vectors $t \in \R^k$ in $\mathcal{U}_{\C}$ and $\mathcal{D}_{\C}$ by $\mathcal{U}$ and $\mathcal{D}$, respectively.
With these, we can express the key message of this work: the local multiplicity $\text{mult}_0 h$ is the birth quota of the original non-generic degeneracy point, i.e, $\text{mult}_0 h$ is the upper bound of the number of Weyl points born from the original degeneracy point:
\begin{equation}
\label{eq:birthquota}
    \sharp h^{-1}_t(0) \leq \text{mult}_0 h.
\end{equation}
Here, $h_t$ is restricted to $B^3_\epsilon$ and $t \in \mathcal{U} \setminus \mathcal{D}$.
Equation \eqref{eq:birthquota} is a consequence of the previous paragraph, and the fact that a real Weyl point is also a complex Weyl point.

\section{Methods to calculate the birth quota}
We mention three different methods to compute the birth quota $\mbox{mult}_0 h_{\C}$. The first one is the direct application of the definition for the \emph{constant deformation} $h_{\C, q}(x)=h_{\C}-q$ with $q \in \C^3$ close to 0. Note that by Sard's lemma \cite{milnortop}, for almost all values $q$, all degeneracy points of $h_{\C, q}$ are generic. By definition, the number of the roots of $h_{\C, q}$ inside the ball $B^6_{\epsilon} $ is the local multiplicity. Hence the computation of the birth quota $\text{mult}_0 h_{\C}$ reduces to solving the equation $h_{\C}=q$, which, in general, can be done numerically.

Another method is based on the fact that the local multiplicity is equal to the local degree $\deg_0 h_{\C}$ of the complexification $h_{\C}$, see \cite[E.3]{mond-ballesteros}. Here, the local degree is understood in the sense described for the minimal model, i.e. it is the local degree of the real-imaginary decomposition (realification) $h_{\C, \R}: \R^6 \to \R^6$. The computation of the local degree also allows the use of integral formulas, see e.g. in \cite{botttu, Asboth}.

We note that an alternative, commonly used definition of the local multiplicity uses algebraic methods: the local multiplicity is defined as the dimension of the so-called \emph{local algebra} of $h$ at $0$, see Appendix~\ref{app:examples} and Refs.~\cite{arnold, mond-ballesteros}. This definition also provides a practical computational method of the birth quota. In the physical applications below, we computed the birth quotas using deformations, as well as this algebraic method.

\section{Applications}
In what follows, we derive the local multiplicity for two different families of isolated twofold degeneracy points: for chiral symmetric band-structure models of few-layer graphene, and for all stable isolated twofold degeneracy points that arise in time-reversal-symmetric crystals. The results are summarised in Table \ref{tab:multiplicities}.

For bilayer graphene, discussed above, we find that the local multiplicity of the unperturbed effective Hamiltonian map $h$ is 4, confirming that the number of newborn 2D Weyl points upon a generic deformation is either 2 or 4.
For chiral-symmetric models of trilayer graphene with ABC stacking, or multi-layer ($n$-layer) graphene with ABCA... stacking, the local degree of the degeneracy point at the $K$ point is $n$, whereas the local multiplicity is $n^2$, as shown in the top panel of Table \ref{tab:multiplicities}. (For derivation, see Example 2 in Appendix~\ref{app:examples}.)

Our methods provide the birth quota for non-generic degeneracy points appearing in 3D crystals as well. 
We focus on time-reversal symmetric crystals, which are classified in 230 space groups.
It is known \cite{ZhiMingYu} that there are four types of isolated twofold degeneracy points in the quasiparticle band structures of such crystals, listed in the bottom panel of Table \ref{tab:multiplicities}. 

The charge-1 Weyl point is the generic degeneracy point (called Weyl point throughout this paper), whereas the charge-2, -3, -4 Weyl points are non-generic degeneracy points showing nonlinear dispersion in certain directions. 
Charge-2 and -3 Weyl points have been proposed in Ref.~\cite{ChenFang_multiweyl}, with effective Hamiltonians in the form $H = a k_z \sigma_z + \left(b k_+^n + c k_-^n\right) \sigma_+ + \textnormal{h.c.}$ with $n \in \{2,3\}$, $|b| \neq |c|$, $k_{\pm} = k_x \pm i k_y$ and $\sigma_{\pm} = \sigma_x \pm i \sigma_y$.
A material proposed to host a charge-2 Weyl point in its electronic band structure in the vicinity of the Fermi energy is SrSi$_2$ \cite{Singh}.
The charge-4 Weyl point has been proposed in Ref.~\cite{TiantianZhang}, and is described by the effective Hamiltonian 
$H=A k_x k_y k_z \sigma_z + B \left(k_x^2+\omega k_y^2+\omega^2k_z^2\right) \sigma_+ + \textnormal{h.c.}$ with $\omega=\exp(-2\pi i/3)$.
Note that this map corresponds to the map studied in Ref.~\cite[Pg. 24]{EisLev}, and also that such charge-4 Weyl points have been observed experimentally \cite{LiLuo,QiaoluChen}.

We have computed the local multiplicities of these degeneracy points, and list the results in Table \ref{tab:multiplicities}.
(Derivations are shown in Example 2 and Example 3 in Appendix~\ref{app:examples}.)
The table also indicates the number of space groups where the corresponding non-generic degeneracy points are stabilised by symmetries.

\begin{table}
	\begin{tabular}{|l|c|c|}
		\hline 
		\thead{$\mathbb{R}^2 \to \mathbb{R}^2$} & \thead{local degree\\ (winding number,\\ top.~charge)} & \thead{local\\ multiplicity\\ (birth quota)} \\
		\hline
		monolayer graphene & 1 & 1 \\
		\hline
		bilayer graphene & 2 & 4 \\
		\hline
		$n$-layer graphene (ABCA...) & $n$ & $n^2$ \\
		\hline
	\end{tabular}
	\begin{tabular}{|l|c|c|c|c|}
	    \hline
		\thead{$\mathbb{R}^3 \to \mathbb{R}^3$} & \thead{local degree\\ (Chern number, \\top.~charge)} & \thead{local\\ multiplicity\\ (birth quota)} & \thead{\#SG\\ SO} & \thead{\#SG\\ nSO}\\
		\hline
		Charge-1 Weyl point & 1 & 1 &  & \\
		\hline
		Charge-2 Weyl point & 2 & 4 & 36 & 48\\
		\hline
		Charge-3 Weyl point & 3 & 9 & 26 & 12\\
		\hline
		Charge-4 Weyl point & $4$ & $12$ & 0 & 13\\
		\hline
	\end{tabular}
	\caption{Local multiplicities of isolated twofold degeneracy points. 
	The upper panel ($\mathbb{R}^2 \to \mathbb{R}^2$) lists the absolute value of the local degree, and the local multiplicity, associated to the electronic quasiparticles that emerge in the simplest chiral-symmetric tight-binding models of mono- or multilayer graphene. The lower panel ($\mathbb{R}^3 \to \mathbb{R}^3$) lists the same invariants, for all four types of isolated two-fold degeneracy points in crystals. The absolute value of the local degree is the minimum number of newborn Weyl points upon a generic deformation. The local multiplicity is the birth quota, i.e., the maximum number of newborn Weyl points.
	The last two columns of the lower panel indicate the number of space groups where symmetries stabilise non-generic degeneracy points, based on Ref.~\cite{ZhiMingYu}. 
	The label SO (nSO) denotes band structures with (without) spin-orbit coupling, e.g. electronic (phononic, photonic) band structures. \label{tab:multiplicities}}
\end{table}

\section{Discussion: physical relevance of the birth quota}

\emph{Band structures for fermions and bosons.}
The local multiplicity of a non-generic twofold degeneracy point in a band structure imposes an upper bound of the number of newborn charge-1 Weyl points from the original degeneracy point upon a perturbation. There are physical setups where this birth of Weyl points has no observable consequence; for example, if the non-generic degeneracy point is `hidden' in an electronic band structure of a crystalline solid, i.e., it is energetically located well below or well above the electrons' Fermi energy (or, at finite temperature, the chemical potential).

However, there are relevant cases where the degeneracy points are in the vicinity of the Fermi energy, and hence the birth of Weyl points upon a perturbation has observable effects. 
Bilayer graphene, and, more generally, $n$-layer graphene with ABCA... stacking \cite{HongkiMin,FanZhang}, are examples where the non-generic degeneracy point coincides with the Fermi energy at zero doping. Also, as noted above, a three-dimensional crystalline material that has been proposed to host a charge-2 Weyl point in the vicinity of the Fermi energy is SrSi$_2$ \cite{Singh}, and processes involving splitting this node have also been studied theoretically~\cite{Naselli2024}.

Weyl nodes and non-generic twofold degeneracy points exist also in bosonic -- e.g., phononic, photonic, magnonic -- band structures, including actual materials as well as metamaterials. 
The physics of degeneracy points is experimentally accessible, as demonstrated, e.g., in the recent works \cite{YihaoYang,HailongHe,QiaoluChen,LiLuo}.

\emph{Perturbations beyond the band-structure framework.}
Certain perturbations, such as mechanical strain, or a Zeeman or exchange field, are taken into account in the non-interacting band structure. 
The birth quota of a non-generic degeneracy point is a very general constraint, relevant for any of such perturbations. For further typical perturbations such as the orbital effect of an external magnetic field, finite sample size, disorder, electron-electron interaction, thermal effects, the constraint imposed by the birth quota is also relevant indirectly. This is because the treatment of the latter perturbations is very often based on the band structure, hence the birth quota constraining the latter has an indirect influence in all further effects based on the band-structure description.

For example, a standard description of the orbital effect of the magnetic field is the minimal coupling in envelope function theory. There, the wave number components of the momentum-space Hamiltonian are promoted to canonical momentum operators, and the vector potential components are added to yield the kinetic momentum operators. The birth quota applies to the starting point of this construction, i.e., it constrains the momentum-space Hamiltonian.
In a similar fashion, the birth quota is also relevant for the further above-listed perturbations that go beyond the non-interacting band structure. 

\emph{Surface states of a 3D topological insulator.}
Surface states of 3D time-reversal invariant topological insulators (3DTIs) exhibit energy degeneracy points in their 2D surface Brillouin zone \cite{HasanRMP2010}.
The energies of these degeneracy points are often located in the bulk band gap.
However, these degeneracies are not robust in the sense that finite-size effects (e.g., a finite height of the slab of the 3D material) break their degeneracy \cite{Pertsova}.
Do the results of our work apply to these 2D band structures?
No; we list multiple reasons for this in what follows.

(i) 
Our results apply when the momentum-dependent effective Hamiltonian map is $\mathbb{R}^n \to \mathbb{R}^n$, i.e., when the dimension of the Brillouin zone is the same as the dimension of the space of traceless effective Hamiltonian matrices.
The surface states of a 3DTI are parametrized by a 2D surface Brillouin zone.
In this case, chiral symmetry would be required to ensure that the effective Hamiltonian map describing the momentum-space vicinity of a zero-energy degeneracy simplifies to $\mathbb{R}^2 \to \R^2$, instead of the general form that is $\R^2 \to \R^3$.
However, chiral symmetry is absent for a generic 2D surface state on the surface of a 3DTI. 
Hence the considerations in this work do not apply for this case.

(ii)  When considering a finite-height slab of a 3DTI, the low-energy subspace of the two surface degeneracy points might become four-dimensional, when the two surface states are at the same energy. In this case, the effective Hamiltonian becomes a map from the 2D surface Brillouin-zone to the space of $4 \times 4$ Hermitian matrices (with some symmetry constraints from TRS), which is not equivalent to an $\R^2 \to \R^2$ map required for our results to be applicable. 

(iii) For a 2D slab of a 3DTI, each band is Kramers degenerate at each time-reversal invariant momentum (TRIM) (e.g., in the origin of the Brillouin zone), and the dispersion relations are linear in the vicinities of these momenta. However, these degeneracy points cannot move in the Brillouin zone upon changing the secondary parameters, if time-reversal symmetry is respected by the perturbations. The presence of a twofold degeneracy point at a TRIM is enforced by time-reversal symmetry, while a degeneracy point away from a TRIM is unprotected and can be gapped out even by time-reversal-symmetric perturbations. 
Hence, the central question of our work -- how many twofold degeneracy points can be born from a non-generic two-fold degeneracy point? -- is not applicable in this case.
It is an interesting question to generalize our results to the case of time-reversal symmetry (and other symmetries) and for band structures with more than two bands. This is, however, beyond the scope of this work.

\emph{Parameter-dependent quantum systems and beyond.}
Weyl points are often described in the context of band structures of crystalline materials or meta-materials. 
However, these Weyl points are special examples of a much broader zoo of robust spectral degeneracy points of parameter-dependent matrices.  
As described by von Neumann and Wigner \cite{Neumann}, Hermitian matrices parametrized by three parameters typically have robust twofold degeneracy points in their parameter space.
(For a more recent description, see, e.g., section II. of \cite{Guba}.)
Examples include Weyl Josephson circuits with at least 3 tunable parameters (e.g., magnetic fluxes, gate voltages) \cite{Riwar,Fatemi,FrankTeleportation,FrankSingularity} and spin systems \cite{Bruno,Scherubl,Frank,Stenger,FrankDensity} controlled by the 3 components of a homogeneous magnetic field.

We refer the interested reader to the following works in particular:
Ref.~\cite{FrankSingularity} provides detailed numerical results describing Weyl points and non-generic degeneracy points in Weyl Josephson circuits; 
Ref.~\cite{Frank} analyses Weyl points and non-generic degeneracy points in the magnetic-field parameter space of a spin-orbit coupled interacting two-spin system.

Von Neumann and Wigner \cite{Neumann} extended their considerations to real symmetric matrices as well, finding that such matrices controlled by two parameters typically have robust twofold degeneracy points in their parameter space.
Exploiting this fact for dynamical matrices of coupled linear oscillators, Ref.~\cite{Guba} exemplifies generic (Weyl-point-like) and non-generic degeneracy points of such classical mechanical systems.
In all the above examples, the concept of birth quota is applicable, and predicts an upper bound of each non-generic degeneracy point exemplified in the above references.

\section{Conclusions}
We have shown that any isolated two-fold non-generic degeneracy point in a 3D configuration space has a birth quota, a maximum number of Weyl points that can be born from the degeneracy point. 
This birth quota can be computed as the local multiplicity of the effective Hamiltonian map associated to the degeneracy point.
We also extended our result to the 2D case with chiral symmetry. 
We have computed the birth quota for the isolated twofold degeneracy points of quasiparticles in 3D crystals, and for the $K$-point electronic degeneracies of multilayer graphene.
An important open question triggered is whether 3-fold or higher-fold isolated degeneracy points exhibit a similar birth quota effect. 
Our result establishes a strong connection between singularity theory and topological semimetals, and also applies to phononic, magnonic, and photonic band structures \cite{HailongHe,YihaoYang,QiaoluChen,LiLuo}.
Furthermore, the concepts and methods discussed here are applicable to parameter-dependent quantum systems, e.g. multi-terminal Josephson circuits \cite{Fatemi,FrankTeleportation,FrankSingularity} or magnetically controlled quantum spin systems \cite{Bruno,Wernsdorfer,Frank}, but also more generally, to physical systems described by matrices, e.g., linearly coupled mechanical oscillators \cite{Guba}, or linear electronic circuits.

\acknowledgments
We acknowledge helpful discussions and correspondence with J. Asb\'oth, A. Bernevig, L. Feh\'er, R. Gim\'enez Conejero, M. Hirschmann, L. Oroszl\'any, and A. Schnyder.
This research was supported by the Ministry of Innovation and Technology (MIT) and the National Research, Development and Innovation Office (NKFIH) within the Quantum Information National Laboratory of Hungary and the Quantum Technology National Excellence Program (Project No. 2017-1.2.1-NKP-2017-00001), by the 
NKFIH fund TKP2020 IES (Grant No. BME-IE-NAT) under the auspices of the MIT, by the NKFIH through the OTKA Grants FK 124723 and FK 132146, and by the EU Horizon Europe grants IGNITE and ONCHIPS.
D.~V. was supported by the Swedish Research Council (VR) and the Knut and Alice Wallenberg Foundation.
D.V. acknowledges support from the Deutsche Forschungsgemeinschaft (DFG, German Research Foundation) under Germany’s Excellence Strategy through the Würzburg-Dresden Cluster of Excellence on Complexity and Topology in Quantum Matter – ct.qmat (EXC 2147, project-ids 390858490 and 392019), and from the NKFIH under OTKA grant no. FK 146499. This project has received funding from the HUN-REN Hungarian Research Network.


\appendix

\section*{Appendix}

In this Appendix, we provide the mathematical background of the results presented in the main text. 
In the main text, the concept of local multiplicity of a real or a complex map was introduced with a perturbative approach; we provide a detailed description of this perturbative approach in Sec.~\ref{app:mult}.
On the other hand, often the local multiplicity is defined with a more direct, algebraic approach, which (i) does not rely on deformations (that is, perturbations), and (ii) provides an alternative computational method. 
We summarize this algebraic description of the local multiplicity and its relation with our approach (without proof) in Sec.~\ref{app:alg}. In Sec.~\ref{app:examples}, we compute the local multiplicity for various examples.

\section{The local multiplicity}\label{app:mult}

In this Appendix, we first prove that in the complex case, the discriminant set does not divide the parameter space into different regions (Sec.~\ref{app:mult1}). 
This statement allows us to define the local multiplicity as the number of complex Weyl points of any generic deformation. 
Then, we prove the conservation of the local degree, 
which, if applied to the complexification, 
implies that the local degree of the complexification is equal to the local multiplicity (Sec.~\ref{app:mult2}). 
Moreover, it provides a method, which we call `constant deformation', to compute the local degree and the local multiplicity (Sec.~\ref{app:mult2}).

\subsection{The local multiplicity is well defined}\label{app:mult1}

Consider a holomorphic map $g: \C^n \to \C^n $, defined on a neighborhood of the origin, with an isolated root at $0$. 
Such a map is called \emph{finite}. 
Let the holomorphic map $\mathcal{G}: \C^n \times \C^k \to \C^n \times \C^k$ be a $k$-parameter unfolding of $g$, i.e. $\mathcal{G}(x, t)=(g_t(x), t)$ and $g_0=g$. The one variable holomorphic function $x \mapsto g_t(x)$ is called the analytic deformation, or simply deformation of $g$ with respect to the control parameter $t$. Note that we use the terminology `deformation' as a more specified version of the ambiguous term `perturbation'. We refer to the two variable function $(x, t) \mapsto g_t(x)$ as  $\mathcal{G}_1(x, t)$, i.e. the first component of $\mathcal{G}$.

As it is explained in the main text for $n=3$, we restrict $\mathcal{G}$ to the domain $B^{2n}_{\epsilon}  \times \mathcal{U}_\C \subset \C^n \times \C^k$ satisfying the following conditions: 
\begin{itemize}

\item $B^{2n}_{\epsilon} \subset \C^n$ is a closed ball centered at the origin with a radius $\epsilon$ small enough that the only root of $g$ in  $B^{2n}_{\epsilon}$ is the origin.

\item $\mathcal{U}_\C \subset \C^k$ is an open ball around the origin such that for all $t \in \mathcal{U}$ the deformation $g_t$ has no roots in the boundary sphere $\partial B^{2n}_{\epsilon}=S^{2n-1}_{\epsilon}$, i.e. $g_t^{-1}(0) \cap S^{2n-1}_{\epsilon}= \emptyset$.

\end{itemize}

As a consequence of these conditions, if $t$ varies in $\mathcal{U}$ along a continuous curve through the origin, then the continuous trajectories of the points of $g_t^{-1}(0)$ do not reach the separator sphere $S^{2n-1}_{\epsilon}$. 
Hence the points of $g_t^{-1}(0) \cap  B^{2n}_{\epsilon}$ can be interpreted as the roots of the deformation born from the origin.

Define the discriminant set $\mathcal{D}_\C \subset \mathcal{U}_\C$ as the set of those $t \in \mathcal{U}_\C$ control parameter values for which $g_t$ has a singular root in $B^{2n}_{\epsilon}$, i.e. there exists $x \in B^{2n}_{\epsilon}$ with $g_t(x)=0$ and $\det (\mbox{Jac}_x (g_t))=0$.

\begin{figure}
	\begin{center}
		\includegraphics[width=0.7\columnwidth]{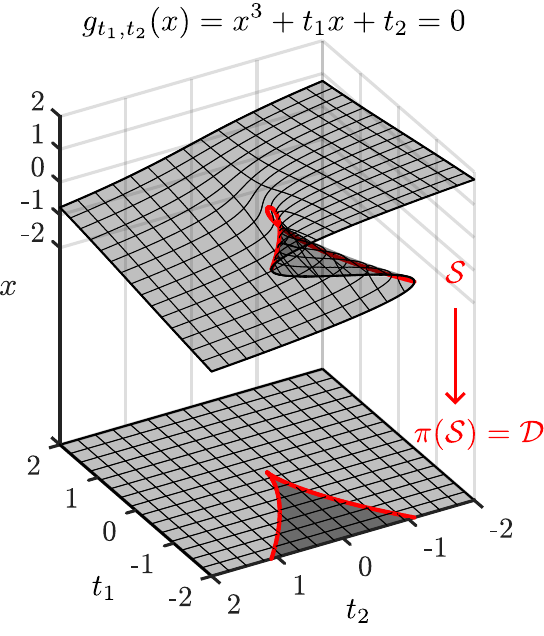}
	\end{center}
	\caption{The projection $\pi$ in the case of the 
	unfolding $\mathcal{G}_1(x,t_1,t_2) = x^3 + t_1 x + t_2$ of $g(x) = x^3$.
	The surface consists of those points $(x,t_1,t_2)$ that satisfy $g_{t_1,t_2}(x) = 0$.
	The set $\mathcal{S}$ is the `contour' of the surface, i.e., the points of the surface where the tangent plane contains a vertical line.
	The projection of $\mathcal{S}$ to the $(t_1,t_2)$ plane is the discriminant set $\mathcal{D}_\C$.
	This discriminant set is the same as the zero locus of the discriminant of the degree-3 polynomial $\mathcal{G}_1$, cf. main text. Note that in this `real picture' the discriminant set divides the $(t_1, t_2)$ plane into two parts such that each point of the lighter part has 1 preimage, while each point of the darker part has 3 preimages with respect to the projection. This is not the case in the complex version, where the complement of the discriminant set is connected and every point of it has 3 preimages.
	\label{fig:cusp}}
\end{figure}

Note that the terminology used here is not universal. The name  `discriminant' usually  denotes the set of the singular values of a map, see \cite{mond-ballesteros}. $\mathcal{D}_\C$ is reminiscent of the `bifurcation set' of an unfolding, but it is not that. More detailed  analysis shows that $\mathcal{D}_\C$ coincides with the bifurcation set with respect to contact equivalence. 
In this work, we call it discriminant set, as it is a straightforward generalization of the discriminant of polynomials in sense described in Section~\ref{s:minmod} and on Figure~\ref{fig:cusp}.

It can happen in special cases that $\mathcal{D}_{\mathbb{C}} = \mathcal{U}_{\mathbb{C}}$, consider for example the unfolding $\mathcal{G}(x, t) =(x^3 -tx^4, t)$ of \mbox{$g(x)=x^3$}. However for generic unfoldings $\mathcal{D}_{\mathbb{C}} \neq \mathcal{U}_{\mathbb{C}}$, that is, there is at least one control parameter $t$ such thet $g(t)$ is a generic deformation. From now on all the unfoldings are assumed to be generic in this sense.

To make the multiplicity well defined, a required property of $\mathcal{D}_\C$ is the following:


\emph{Proposition 1.} 
For a finite map $g$, the  complement $\mathcal{U}_\C \setminus \mathcal{D}_\C$ of $\mathcal{D}_\C$ is path connected: any two parameter values $t_1, t_2 \in \mathcal{U}_\C \setminus \mathcal{D}_\C$ can be joined with a continuous curve in $\mathcal{U}_\C \setminus \mathcal{D}_\C$.

\emph{Proof:} We have not found this statement in the literature, hence we provide a short proof based on two facts.

 (1) A set defined by holomorphic equations in the complex space is called complex analytic set. A complex analytic set has at least 1 complex codimension, hence its real codimension is at least 2, therefore its complement is path connected \cite{dejong,loja}. 
 
 (2) Remmert's Finite Mapping Theorem \cite[Chapter V]{loja}: The image of a finite map between complex analytic sets is complex analytic. 
 
By (1) it is enough to prove that $\mathcal{D}_\C$ is an analytic set in $\mathcal{U}_\C$. To prove it we use (2).

Observe that $\det (\mbox{Jac}_x (g_t))=0$ holds at a point $(x, t) \in B^{2n}_{\epsilon}  \times \mathcal{U}_\C$ if and only if $\det (\mbox{Jac}_{(x,t)} (\mathcal{G}))=0$. Therefore $t \in \mathcal{D}_\C$ if and only if $ (0, t) $ is a singular value of $\mathcal{G}$, that is, there exists $x \in B^{2n}_{\epsilon} $ such that $\mathcal{G}(x, t)=(g_t(x), t)=(0, t)$ and $\det (\mbox{Jac}_{(x,t)} (\mathcal{G}))=0$. 

To formalize the former description, we define 
\bean
\mathcal{S}=
\{(x, t) \in  B^{2n}_{\epsilon} \times \mathcal{U}_\C \ | && \ 
\det (\mbox{Jac}_{(x,t)} (\mathcal{G}))=0,
\\
\nonumber
&& 
\mathcal{G}_1(x,t)=0
\}.
\eean
$\mathcal{S}$ is a complex analytic set.
Consider the projection map $\pi: \mathcal{S} \to \C^k$, $\pi(x, t)=t$; for a specific example and visualization, see Fig.~\ref{fig:cusp}. 
Then, since $g$ is finite, the projection map $\pi$ is finite from $(\mathcal{S}, 0)$ to $(\C^k, 0)$ in the sense that 0 is an isolated point of $\pi^{-1}(0)$. By  Remmert's Finite Mapping Theorem, the image of $\pi|_{\mathcal{S}}$ is complex analytic. 
On the other hand, the image $\pi(\mathcal{S})$ is $\mathcal{D}_\C$.
Therefore $\mathcal{D}_\C$ is complex analytic.
$\blacksquare$

As a consequence of Proposition 1, the cardinality of the set $g_t^{-1}(0) \cap B^{2n}_{\epsilon}$ is the same for any $t \in \mathcal{U}_\C \setminus \mathcal{D}_\C$. 
We call this number the local multiplicity $\mbox{mult}_0 g$ of $g$. 
We define the local multiplicity $\mbox{mult}_0 h$ of a real analytic map $h: \R^n \to \R^n$ as the local multiplicity $\mbox{mult}_0 h_\C$ of the complexification $h_\C: \C^{n} \to \C^{n}$, where $h_\C $ is the complex holomorphic map defined with the same power series as $h$. 

Note that $\mbox{mult}_0 h$ is well-defined with this definition if and only if $h_\C$ is finite at 0. 
Moreover, the local multiplicity is always a positive integer. 
We already anticipate that the alternative, algebraic definition
of the local multiplicity (see App.~\ref{app:alg}) can be applied to any analytic map (i.e., not only finite maps).
Furthermore, according to the algebraic definition, non-finite analytic maps will have an infinite local multiplicity.

\subsection{The local multiplicity is the local degree of the complexification}
\label{app:mult2}

Consider a point $P \in \R^m$, and an analytic map $h: U_P \to \R^m$ defined on a neighborhood $U_P$ of $ P$ on which the Taylor series of $h$ converges. Assume that $P$ is an isolated root of $h$, i.e., $h(P)=0$, and $P$ is an isolated point of the preimage set $h^{-1}(0)$.

The \emph{local degree} $\deg_P h$ of $h$ at $P$, also called the \emph{index} of $h$, is defined as follows \cite{milnortop, arnold, Eis, EisLev, mond-ballesteros}.
Take a sphere $S^{m-1}_{\epsilon} \subset U_P \subset \R^m$ around $P$ with a small enough radius $\epsilon$ such that $h$ has no other zeros inside the sphere, and the orientation of the sphere is inherited from $\R^m$.
Define the map
\begin{equation}\label{eq:index}
    \widetilde{h}=\frac{h}{|h|}: S^{m-1}_{\epsilon} \to S^{m-1},
\end{equation}
which is called `pseudospin texture' in physics, see e.g. \cite{TiantianZhang}.
Then the local degree $\deg_P h$ of $h$ is defined as the \emph{global degree} $\deg \widetilde{h}$ of $\widetilde{h}$, defined as follows. Take a \emph{regular value} $q \in S^{m-1}$ of $\widetilde{h}$, i.e. the tangent map (the Jacobian) of $\widetilde{h}$ has maximal rank at each preimage $p \in \widetilde{h}^{-1}(q)$. Each preimage $p$ is endowed with a sign, which is positive, if $\widetilde{h}$ preserves the orientation around $p$, and it is negative, if the orientation is reversed. 
Then, $\deg \widetilde{h}$ (and hence $\deg_P h$) is defined as the sum of the signs of the the preimages of a regular value $q$. 

Importantly, the global degree $\deg \widetilde{h}$ does not depend on the choice of $q$, and $\deg_0 h=\deg \widetilde{h}$ does not depend on the choice of $\epsilon$, see \cite[E.3]{mond-ballesteros} and \cite{milnortop}.

A simple example, which is easy to visualise, is the $m=2$ case, when the local degree is also called `winding number'. 
For example, the map 
\bean \label{eq:znegyzet}
h&:& \R^2 \to \R^2, \\
\nonumber
h(x,y) &=& (\mbox{Re} (x+i y)^2, \im(x+i y)^2)=(x^2-y^2, 2xy),
\eean
has local degree $\deg_0 h=2$. In fact, $h$ maps the unit circle of the domain onto the unit circle of the target, and any value $q=(X, Y) \in S^1$ is a regular value, which has two preimages, namely, $x+iy=\pm \sqrt{X+iY}$. 
The map $\widetilde{h} = h|_{S^1}: S^1 \to S^1$ preserves the orientation, hence both preimage has positive sign, implying $\deg_0 h = 2$.

In the case of an $h: \R \to \R$ function, $\deg_0 h$ is either $\pm 1$ or 0 depending on the order of the Taylor series of $h$: if the power of the leading term is odd, then the local degree is $\pm 1$, and if this degree is even, the local degree is $0$.

The local degree $\deg_P g$ of a complex map $g: \C^n \to \C^n $ at an isolated root $P \in \C^n$ is defined as the local degree $\deg_P g_\R$ of the realification (real-imaginary decomposition) $g_\R: \R^{2n} \to \R^{2n}$ of $g$.
For example, consider $g: \C \to \C$, $g(z)=z^2$. Its realification is the map $h$ in Eq.~\eqref{eq:znegyzet}, hence $\deg_0 g=2$. 
Note that in general, the local degree of a real map is not equal to the local degree of its complexification. In fact, take the function $h: \R \to \R$, $h(x)=x^2$, then $\deg_0 h=0$, while $\deg_0 h_\C=\deg_0 h_{\C, \R}=2$.

Consider an unfolding $\mathcal{H}: \R^m \times \R^k \to \R^m \times \R^k$, of $h: \R^m \to \R^m$ where $\mathcal{H}(x, t)=(h_t(x), t)$ and $h_0=h$. Restrict $\mathcal{H}$ to a domain $B^m_{\epsilon} \times \mathcal{U} $ as it is explained in the main text:
first we find a domain $B^{2m}_{\epsilon} \times \mathcal{U}_\C$ for the complexification $h_\C$ (cf. Appendix~\ref{app:mult1}), and then define $\mathcal{U} \subset \R^k $ as the set of the real parts of the points of $\mathcal{U}_\C$, and $B^{m}_{\epsilon}$ as the set of the real parts of the points of $B^{2m}_{\epsilon}$.

\emph{Proposition 2.} 
(\emph{Conservation of the local degree.})
The sum of the local degrees at the roots of a deformation $h_t$ equals the local degree of the original map $h$ at its single root.
Formally:
\bean \label{eq:consdeg}
    \sum_{p \in h_t^{-1}(0) \cap B^m_{\epsilon}} \deg_p h_t &=&\deg_0 h,
\eean
holds for all $t \in \mathcal{U}$.

\emph{Proof:} 

\begin{figure}
	\begin{center}
		\includegraphics[width=0.5\columnwidth]{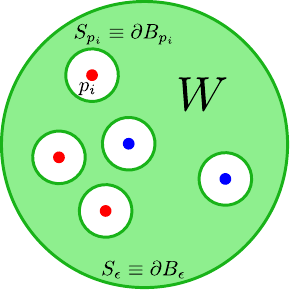}
	\end{center}
	\caption{Visual proof of Proposition 2. The boundary of $W$ consists of the sphere $S_{\epsilon}^{n-1}$ and the spheres $S^{n-1}_p$ around each Weyl point $p \in h_t^{-1} (0) \cap B_{\epsilon}^m$ with reversed orientation. Since the degree of the map $\widetilde{h}_t$ is 0 on the whole boundary of $W$, the degree on $S_{\epsilon}^{n-1}$ is equal to the sum of the signs  of the Weyl points inside the sphere.
	\label{fig:degree}}
\end{figure}

Take disjoint closed balls $B^m_p \subset B^m_{\epsilon}$ around the roots of $h_t$ with boundary $S^{m-1}_p$ and interior $\mbox{int} (B^m_p)$, see Fig.~\ref{fig:degree}. 
Cut out the interior of these disjoint balls from  $B^m_\epsilon$ to obtain the closed domain
 \bean
 W=B^m_{\epsilon} \setminus \bigcup_{p \in h_t^{-1}(0) \cap B^m_{\epsilon}} \mbox{int} (B^m_p).
 \eean
Furthermore, consider the normalized map on $W$, defined as
 \bean
 \widetilde{h}_t=\frac{h_t}{|h_t|}: W \to S^{m-1}.
 \eean
 
 By a classical topological argument, the global degree of the restriction
 \bean
 \widetilde{h}_t|_{\partial W}: \partial W \to S^{m-1} 
 \eean
 to the boundary $\partial W$ of $W$ is 0, see \cite[pg.~28]{milnortop}.
 This global degree is 
 \bean
 \deg \widetilde{h}_t|_{\partial W}=\deg \widetilde{h}_t|_{S^{m-1}_{\epsilon}}-  \sum_p \deg \widetilde{h}_t|_{S^{m-1}_p},
 \eean
where the minus sign arises because the boundary components $S^{m-1}_p$ inherit reversed orientation from $W$. 
Therefore we find 
\bean
\sum_p \deg \widetilde{h}_t|_{S^{m-1}_p}
=
\deg \widetilde{h}_t|_{S^{m-1}_{\epsilon}}.
\eean

The left hand side is equal to the left hand size of Eq.~\eqref{eq:consdeg} by definition. 
The right hand side is equal to $\deg_0 h$, since $\widetilde{h}_t|_{S^{m-1}_{\epsilon}}$ and $\widetilde{h}|_{S^{m-1}_{\epsilon}}$ are homotopic maps, and the global degree is homotopy invariant \cite[pg.~28]{milnortop}.
This concludes the proof of Eq.~\eqref{eq:consdeg}. $\blacksquare$

Now we derive two important corollaries of Proposition 2: (1) An alternative definition of the local degree, which is well suited to calculate the local degree. (2) Local multiplicity is the local degree of the complexification.
 
 As the first step towards (1) and (2), let $\mathcal{D} \subset \mathcal{U}$ be the discriminant set of $\mathcal{H}$, and consider a control parameter vector $t \in \mathcal{U} \setminus \mathcal{D}$. The Jacobian of $h_t$ at each root $p \in B^m_{\epsilon}$ has the maximal rank $m$, therefore $\deg_p h_t= \pm 1$, and its sign is equal to the sign of the determinant of the Jacobian. 
 By Eq.~\eqref{eq:consdeg}, the local degree $\deg_0 h$ is equal to the sum of the signs of the roots $p$ of $h_t$. 

As the second step toward (1) consider a specific simple deformation, a \emph{constant deformation} $h_q=h-q$ with a regular value $q \in \R^m $ close to the origin. 
More precisely, we consider the one-parameter deformation $h_t=h-tq$ on the restricted  domain $\mathcal{U} \subset \R$, with $\mathcal{D}=\{0\}$, and take $t=1$. Then $h_q^{-1}(0)=h^{-1}(q)$. Therefore, $\deg_0 h$ is equal to the sum of the signs of the preimages of a small regular value $q$, i.e.,
\bean
\deg_0 h =  
\sum_{p \in h^{-1}(q) \cap B^m_\epsilon} 
\text{sgn} \left(
    \det \left(
        \text{Jac}_p (h)
    \right)
\right)
.
\eean
 This result can be used to calculate the local degree $\deg_0 h$.
 
 Finally, we derive (2). We apply Proposition 2 for a deformation $g_t$ of a  complex holomorphic map $g: \C^n \to \C^n$, where $t \in \mathcal{U}_\C \setminus \mathcal{D}_\C$.
 A holomorphic map always preserves the orientation, hence $\deg_p g_t=1$ holds for each root $p$ of $g_t$. Therefore the sum at the left side of Eq.~\eqref{eq:consdeg} is the number of the roots, that is, the multiplicity $\mbox{mult}_0 g$. 
 We conclude that 
  \bean \label{eq:degmult}
  \mbox{mult}_0 g=\deg_0 g
  \eean
  holds for finite complex maps.
 
\section{Algebraic approach: the local algebra}\label{app:alg}

In the singularity-theory literature, multiplicity is often defined in an algebraic way, such that the definition refers to the real map, without referring to its complexification \cite{arnold, mond-ballesteros, dejong}. In fact, the multiplicity $\mbox{mult}_0 h$ is the dimension of the local algebra $Q_0(h)$ associated to $h$ at 0. 
In Sec.~\ref{app:algebraintro}, we summarize concepts and relations of elementary algebra (ring theory).
In Sec.~\ref{app:localalgebra}, we introduce the local algebra and use it to provide an alternative definition of the local multiplicity.

\subsection{From a ring to a local algebra}
\label{app:algebraintro}

An elementary example of a ring is $(\Z, +, \cdot)$, the integers equipped with addition and multiplication. The
integers with addition form a group. The integers with multiplication form a semi-group, which has an identity element (that is 1),  but not every integer
is invertible: 0 has no inverse, 1 has an inverse (itself), $-1$ has an inverse (itself), but other integers do not have
their inverse (within $\Z$). In addition, the multiplication is commutative. Actually, the integers form a \emph{commutative ring with an identity element}.

Take, for example, the subset $I_4 = \{\dots, -8,-4,0,4,8, \dots\}$ of the ring of integers.
It contains $0$. 
It is closed under both addition and multiplication.
In addition, it is also closed under `external multiplication'; that is, under multiplication with ring elements -- integers -- that are outside $I_4$.
Such a subset is called an \emph{ideal} of the ring.

A single element of the ring, say $r= 4$, can generate a \emph{principal ideal}, which is $I_r = I_4$ in this case. 
Formally, the principal ideal generated by $r\in \Z$ is $I_r = \{r \cdot r' | r' \in \Z\}$.

Two elements of the ring can also \emph{generate an ideal}: 
$I_{r,q} = \{a \cdot r + b \cdot q | a,b \in \Z\}$.
For example, $I_{2,3} = R = I_1$. 
Another example is $I_{2,4} = I_2$.
The ring of integers is a \emph{principal ideal ring}: every ideal in $\Z$ is a principal ideal.
Furthermore, the ideal generated by a finite set of integers is the ideal generated by their greatest common divisor. 
The ring of integers is a \emph{Noetherian ring}: every ideal in $\Z$ is finitely generated.
(Every principal ideal ring is a Noetherian ring.)

A ring can be partitioned with an ideal, e.g., $\Z/I_4$. 
The elements of the ring are assigned to the same partition class (are considered equivalent) if their difference is  in the ideal: $r_1 \sim_I r_2$ if $r_1-r_2 \in I$.
That is, partitioning $\Z$ with $I_4$ yields 4 classes, the well-known residue classes, $[0]$, $[1]$, $[2]$, $[3]$. 
For example $[3] = \{\dots,-5, -1,3,7,\dots\}$.

The set of classes is called the \emph{quotient} $R/I$ of ring $R$ with respect to its ideal $I$, 
and the elements of the quotient $R/I$ are often called \emph{residue classes} of $R$ $\mod I$.
In fact, the quotient $R/I$ is a ring.
For example, $[2]+[3]=[1]$, in the sense that any representative of $[2]$ plus any representative of $[3]$ is contained in $[1]$.
Similarly, $[2] \cdot [3] = [2]$. 

In certain quotient rings we find pairs which multiply to $[0]$, even though neither of them is $[0]$.
For example, in e.g., in the quotient ring $\Z/I_6$, we have $[2] \cdot [3]=[0]$. 
Such pairs are called zero divisors. 

A commutative ring is called \emph{domain} (or \emph{integral domain}) if it is free of zero divisors. 
That is, if $a \cdot b=0$ implies $a=0$ or $b=0$. 
For example, $\Z$ is  a  domain, but $\Z/I_n$ is a domain only if $n$ is a prime.

A ring is called a \emph{real algebra}, if it is also a vector space over $\R$, and the multiplication with scalars is compatible with the ring multiplication. In the same way complex algebras are rings which are also vector spaces over $\C$. The typical examples are function algebras. In fact, the functions from any set $A$ to $\R$ form a real algebra with respect to the pointwise operations, and if a subset of them is closed to the operations, then it is a real algebra as well. In these algebras the ring multiplication is also commutative, and they have identity element, namely, the constant 1 function.

The simplest functions are the polynomials. $\R[x]$ denotes the real algebra of polynomials of one variable with real coefficients, and similarly $\C[x]$ is the complex algebra of complex polynomials. The pointwise algebra operations agrees with the formal operations using the coefficients, i.e.
\bean\label{eq:sum}
\left( \sum_{i=0}^n a_i x^i \right) +
\left( \sum_{i=0}^m b_i x^i \right)= \sum_{i=0}^k (a_i+b_i) x^i 
\eean
where $k=\max(n, m)$, and
\bean\label{eq:mult}
\left( \sum_{i=0}^n a_i x^i \right) \cdot
\left( \sum_{i=0}^m b_i x^i \right)= \sum_{i=0}^l \left( \sum_{j=0}^{i} a_j b_{i-j} \right) x^i 
\eean
where $l=nm$.

$\R[x]$ and $\C[x]$ are principal ideal rings as well, it follows from the existence of \emph{greatest common divisor} of polynomials. Moreover these rings have another important common property with $\Z$: the unique irreducible decomposition. In $\Z$ the irreducible elements are the prime numbers, and by the fundamental theorem of number theory, every integer except 0, 1 and $-1$, can be decomposed as a product of prime numbers. The prime factors are unique (up to reordering) and their signs. Similarly, in $\C[x]$ the irreducible polynomials are the degree 1 (i.e., linear) polynomials, and by the fundamental theorem of algebra every polynomial of at least degree 1 can be decomposed as the product of degree 1 polynomials. The irreducible factors are unique (up to reordering and up to scalar factors) -- in fact, in $\C[x]$ the invertible elements are the degree 0 polynomials (i.e. constant polynomials). Referring to this property we say that $\Z$ and $\C[x]$ are \emph{unique factorization domains (UFD)}. $\R[x]$ is also an UFD, the irreducible elements are the degree 1 polynomials and the degree 2 polynomials with negative discriminant. All the real polynomials of degree at least 1 can be decomposed to the product of irreducible ones, and the decomposition is essentially unique.

The multi-variable polynomial algebras $\R[x_1, \dots, x_m]$, and $\C[x_1, \dots, x_m]$ are UFDs as well, but they are not principle ideal rings. In fact, the ideal generated by $x_1$ and $x_2$ cannot be generated by only one element.
Note that a principal ideal domain is UFD, but the converse is not true, $\R[x_1, \dots, x_m]$, and $\C[x_1, \dots, x_m]$ are counterexamples. On the other hand these algebras are Noetherian by the Hilbert's basis theorem. 

Further examples are the algebras of \emph{formal power series} $\R[[x_1, \dots, x_m]]$ and $\C[[x_1, \dots, x_m]]$ of $m$ variables. E.g. $\R[[x, y]]$ consists of the infinite power series
\bean
g(x, y)=\sum_{i, j=0}^{\infty} a_{ij} x^i y^j,
\eean
where $a_{ij} \in \R$.
Since these power series are not assumed to be convergent, they are not functions. 
Nevertheless, the ring operations are defined formally, using the coefficients, similarly to Eq.~\eqref{eq:sum} and Eq.~\eqref{eq:mult}.

The rings $\R[[x_1, \dots, x_m]]$ and $\C[[x_1, \dots, x_m]]$ are Noetherian UFDs as well. Moreover they have an important additional property: they are \emph{local} rings. Consider the power series $g$ with constant term $g(0)$ not zero. These are exactly the invertible elements. Indeed, the multiplicative inverse of  $g=g(0)-\widetilde{g}$ with nonzero $g(0)$ reads
\bean
g^{-1}=\frac{1}{g(0)} \cdot \left(1-\frac{\widetilde{g}}{g(0)}\right)^{-1},
\eean
which can be rewritten in the form of a formal power series using the geometric series formula as
\bean
g^{-1}=
\frac{1}{g(0)} \cdot \sum_{k=0}^{\infty}
\left(\frac{\widetilde{g}}{g(0)}\right)^{k}.
\eean
However, a power series with $g(0)=0$ does not have a multiplicative inverse. 

On the other hand, the collection of power series with $g(0)=0$ form an ideal $\mathfrak{m}_m $. 
This $\mathfrak{m}_m $ is a \emph{maximal ideal}, since the only ideal containing it is the whole ring. Moreover, $\mathfrak{m}_m $ is the only maximal ideal in $\R[[x_1, \dots, x_m]]$ (resp. in $\C[[x_1, \dots, x_m]]$). A ring (resp.~an algebra) with a unique maximal ideal is called \emph{local ring} (resp.~ \emph{local algebra}). Being local is equivalent with the fact that the complement of the invertible elements forms an ideal. 

Finally, the algebra in which we will work is $\R\{x_1, \dots, x_m\} \subset \R[[x_1, \dots, x_m]]$ (resp. $\C\{x_1, \dots, x_m\} \subset \C[[x_1, \dots, x_m]]$), consisting of the power series with nonzero convergence radius. Such a power series defines an analytic function $g: \R^m \supset U_0 \to \R$ on a neighborhood $U_0$ of the origin. Hence $\R\{x_1, \dots, x_m\}$ (resp. $\C\{x_1, \dots, x_m\} $) can be interpreted as either as the algebra of the locally convergent power series defined in the neighborhood of zero, or as the algebra of the analytic functions defined in a neighborhood of zero. 
These algebras are Noetherian UFD local algebras as well.

\subsection{Local algebra of analytic maps}  
\label{app:localalgebra}

In what follows, we will consider $f: \R^m \supset U_0 \to \R^m $ as an analytic map defined in a neighborhood $U_0$ of the origin with $f(0)=0$, simply denoted by \mbox{$f:(\R^m, 0) \to (\R^m, 0)$}. We will denote the components as $f_j: U_0 \to \R$ with $j \in \{1, \dots, m\}$. 
Note that in the main text we considered the special cases $1 \leq m  \leq 3$.

Let $\R\{x_i\}$ ($i=1, \dots, m$) denote the set of locally convergent power series of $m$ variables, i.e. $\R^m \to \R$ functions that are analytic at the origin. 
Let \mbox{$I_f=I(f_1, \dots, f_m)$} be the ideal generated by the components $f_j$ of $f$. Recall that $I_f$ consists of all linear
combinations of $f_j$ with coefficients in $\R\{x_i\}$. The \emph{local algebra} of $f$ is the quotient $Q_0(f)=\R\{x_i\}/I_f$, i.e. the algebra of the residue classes $\mod I_f$.

The local algebra $Q_0(f)$ of any power series $f$ (defined in the previous paragraph) is a local algebra (defined in the previous subsection). 
Indeed, the residue classes of the elements of $\mathfrak{m}$ form the unique maximal ideal of it.

In this algebraic framework, the \emph{local multiplicity} $\mbox{mult}_0 f$ of $f$ is defined as the dimension $\dim Q_0(f)$ of the local algebra, cf. \cite[Prop. 2.2.]{EisLev}. 

\emph{Example 1:} Take $f_1(x, y)=x$ and $f_2(x, y)=y^2$. 
Then, the generated ideal in $\R\{x, y\}$ is \bean
I_f=\{ &a&_1(x, y) \cdot x +a_2(x, y) \cdot y^2 \ |
\\ \ &a&_1, a_2 \in \R \{x, y \} \},\nonumber
\eean
i.e. $g \in I_f$ if and only if every term of $g$ is divisible by $x$ or $y^2$. The local algebra of $f(x, y)=(x, y^2)$ is formed by the linear combinations $a[1]+b[y]$ ($a, b \in \R)$ of the residue classes of $[1]$ of $1$ and $[y]$ of $y$. Indeed, $g_1$ and $g_2$ in $\R\{x, y\}$ represent the same residue class if and only if each term of $g_1-g_2$ is divisible by $x$ or $y^2$, consequently the residue class $[g]$ of
\bean
g(x, y)&=&a_{00}+a_{10}x + a_{01}y+\\
&&a_{20}x^2+a_{11}xy + a_{02}y^2 + \dots\nonumber
\eean
is $a_{00} [1]+a_{01}[y]$. 
Therefore, in this example, the local multiplicity $\text{mult}_0 f$ of $f$ is $\text{dim} Q_0(f)=2$.

\emph{Example 2:} The local multiplicity of a single-variable power series $f \in R\{x\}$ is equal to its order $r$, for the following reason:
\bean
f(x)=\sum_{i=r}^{\infty} a_i x^i=x^r \cdot \sum_{j=0}^{\infty} a_{j+r} x^j,
\eean
hence $f$ agrees with its leading term $x^r$ up to an invertible factor. Therefore $I_f=I({x^r})$. A basis of the quotient $Q_0 (f)=\R\{x\}/I_f$ is formed by the residue classes $[1], [x], [x^2], \dots, [x^{r-1}]$.
As a consequence, the local multiplicity of $f$ is indeed $r$.
This example also shows that the local multiplicity is a generalization  of the order of single-variable power series.

In the rest of this section, we connect this algebraic definition of the local multiplicity with the concepts used in the main text and in App.~\ref{app:mult}.

Recall that the complexication of a real analytic map \mbox{$f: (\R^{m}, 0) \to (\R^{m}, 0) $} is the holomorphic map \mbox{$f_\C: (\C^m, 0) \to (\C^m, 0) $} defined by the same power series.
The local algebra $Q_0(f_\C)=\C\{x_i\}/I_{f_\C}$ of $f_\C$ is a complex algebra, and in fact, \mbox{$Q_0(f_\C)\cong Q_0(f) \otimes \C$}. Hence the local  multiplicities $\mbox{mult}_0 f_\C=\dim_\C Q_0(f_\C)$ and $\mbox{mult}_0 f=\dim Q_0(f)$ are equal. 
On the other hand, if $g_\R: (\R^{2m}, 0) \to (\R^{2m}, 0) $ is the realification of a complex map $g: (\C^m, 0) \to (\C^m, 0) $, then $\left(\mbox{mult}_0 g \right)^2=\mbox{mult}_0 g_\R$ holds, see the proof of \cite[Prop. 2.4.]{EisLev}.

A real analytic map $f$ is called \emph{finite} if its local multiplicity \mbox{$\mbox{mult}_0 f=\dim Q_0 (f)$} is finite. This is  is equivalent with the fact that 0 is an isolated root of the complexification $f_\C$, which implies that 0 is an isolated root of $f$ as well, see \cite[Thm. D.5.]{mond-ballesteros}.
For a finite map $f$
the local degree of the complexification is equal to the multiplicity, i.e., $\deg_0 f_\C =  \mbox{mult}_0 f$, see \cite[Cor. E.3]{mond-ballesteros}. 
Together with the arguments in Appendix~\ref{app:mult2} we conclude the equivalence of the two definitions of the local multiplicity. Namely, the number of roots born from the origin upon a generic deformation $f_{\C, t}$ is equal to the dimension $\dim Q_0(f)$ of the local algebra $Q_0(f)$.

Moreover, the local degree $\deg_0 f$ of the real map $f$ can also be computed from the local algebra: it is the index of a suitably defined symmetric bilinear form on $Q_0(f)$, that is, the difference between the number of positive and negative eigenvalues, see \cite{Eis, EisLev}. This also implies $|\deg_0 f| \leq \mbox{mult}_0 f$. These relations are summarized in Table~\ref{tab:table}.

\begin{table*}
	\centering
	\begin{tabular}{|c|c|}
	\hline &\\[2.0ex]
		\underline{Complexification} & \underline{Realification}\\[2.0ex]
		$\begin{array}{c}
		f: (\R^m, 0) \to (\R^m, 0) \ \Rightarrow  \ f_\C: (\C^m, 0) \to (\C^m, 0) \\[2.0ex]
        \mbox{mult}_0 f \overset{1.}{=} \mbox{mult}_0 f_\C \overset{2.}{=} \deg_0 f_\C,\\[2.0ex]
        |\deg_0 f| \overset{3.}{\leq} \deg_0 f_\C 
		\end{array}$
		&
		$\begin{array}{c}
		g: (\C^n, 0) \to (\C^n, 0)\ \Rightarrow  \ g_\R: (\R^{2n}, 0) \to (\R^{2n}, 0)\\[2.0ex]
        \mbox{mult}_0 g \overset{4.}{=} \deg_0 g \overset{5.}{=} \deg_0 g_\R,\\[2.0ex]
        \mbox{mult}_0 g_\R \overset{6.}{=} (\mbox{mult}_0 g)^2 
		\end{array}$\\[2.0ex]
		&\\[2.0ex]
		\hline
	\end{tabular}
	\caption{
	Summary of the relations between local degree and local multiplicity, complexification and realification. 1.: According to the definition of the local multiplicity introduced in the main text, this equation holds by definition. In fact, both $\mbox{mult}_0 f_\C$ and $\mbox{mult}_0 f$ are defined  as the number of the generic roots of a generic deformation of $f_\C$. On the other hand in Appendix~\ref{app:localalgebra} $\mbox{mult}_0 f$ is defined as the real dimension of the local algebra $Q_0(f)$ of $f$ and $\mbox{mult}_0 f_\C$ is defined as the complex dimension of the local algebra $Q_0(f_\C)$ of $f_\C$. Based on this, the equation follows from the fact that $Q_0(f_\C)$ is the complexification of $Q_0(f)$. The equivalence of the two different definitions of the local multiplicity is not proved in this article, we refer to \cite{mond-ballesteros}. 2.: The equation is proved in Appendix~\ref{app:mult2}, see Eq.~\eqref{eq:degmult}. More precisely the equation is proved for that definition of the local multiplicity we introduced in the main text. 3.: The inequality is proved in the main text (using the definitions introduced there). For singular germs, i.e. $\mbox{Jac}_0 (f) <m $, the sharper inequality $2|\deg_0 f| \leq \mbox{mult}_0 f$ is proved in \cite{EisLev}. 4.: This is equation  Eq.~\eqref{eq:degmult} in Appendix~\ref{app:mult2}. 5.: The equation holds by definition. 6.: The equation is proved in \cite{EisLev}. \label{tab:table}}
\end{table*}

\section{Computing the local degree and the local multiplicity for a few examples}\label{app:examples}

In the following section we compute the above invariants on several examples using the introduced methods. Since most of them are motivated by physical systems, we use the physical and the mathematical terminology as synonyms. Namely, the roots of a map  \mbox{$h:(\R^3, 0) \to (\R^3, 0)$} or a deformation $h_t$ of $h$ are called degeneracy points, the generic roots of $h_t$  are the Weyl points. The local degree of each root is its topological charge, and the local multiplicity $\mbox{mult}_0 h$ is the birth quota.

\subsection*{Example 1}

Consider the effective Hamiltonian map
\bean
h(x, y, z) = 
\begin{pmatrix}
x\\
y\\
z^n
\end{pmatrix}.
\eean
The case $n=3$ is described in \cite{Konye} as the three-node process, where two Weyl points with the same charge collide with an oppositely charged third one. This suggests that the charge (local degree) is +1 and the local multiplicity is 3.

\emph{Local degree from the constant-deformation method.}
For a general $n$, $\deg_0 h= n\text{ mod }2$. Indeed, the constant deformation
\bean\label{eq:nthorderpert}
h_t(x, y, z)=h(x,y,z)-(t_1, t_2, t_3),\hspace{6mm} t_3 < 0
\eean
has 1 real root
\bean\label{eq:nthorderpert2}
(x,y,z)_\text{WP}=\left(t_1,t_2,\sqrt[n]{t_3}\right)
\eean
for $n$ odd with positive sign (since $z^n$ is increasing), and there is no solution for even $n$. Note that geometrically, $\deg_0 h$ is the `generalized winding number' of the restricted normalized map $\frac{h}{|h|}|_{S^{2}}: S^2 \to S^2$.

\emph{Local multiplicity from the constant-deformation method.}
Here, we show that $\mbox{mult}_0 h=n$. Indeed,
\bean\label{eq:nthorderpert3}
h_\C(x, y, z)=(t_1, t_2, t_3),\hspace{6mm} t_3 \neq 0
\eean
has $n$ complex solutions
\bean\label{eq:nthorderpert4}
(x,y,z)_{\C\text{WP},k}=\left(t_1,t_2,e^{i\frac{2\pi k}{n}}\sqrt[n]{t_3}\right),\hspace{6mm}
\eean
with $0\leq k < n$, hence $\deg_0 h_\C=n$, and this is equal to $\mbox{mult}_0 h_\C$ which is equal to $\mbox{mult}_0 h$. Geometrically, $n$ is the `generalized winding number' of the restricted normalized map $\frac{h_\C}{|h_\C|}|_{S^{5}}: S^5 \to S^5$.

\emph{Local multiplicity from the algebraic method.}
We present here the algebraic method step by step, to illustrate it on this simple example. The ideal $I_h \subset \R\{x, y, z \}$ consists of the (locally convergent) power series of 3 variables such that every term is a multiple of $x$, $y$ or $z^n$. Two elements $f$ and $g$ of $\R\{x, y, z \}$ represents the same residue class $\mod I_h$ if and only if $f-g \in I_h$, that is, the coefficients of the $z^k$ terms are the same in $f$ and $g$ for $k<n$. In other words, the element $[f] \in \R\{x, y, z \} /I_h$ can be represented by the power series of $z$ derived from $f$ by substituting $x=y=z^n=0$. We obtained that $Q_0(h)=\R\{x, y, z \} /I_h=\R\{z\}/I{(z^n)}$ is a vector space with a  possible choice of a basis is formed by the residue classes of $1, z, z^2, \dots, z^{n-1}$. We conclude that $\mbox{mult}_0 h=\dim Q_0(h)=n$.

 All the above arguments show that the only essential part of this $h$ is the third component $z^n$. The study of the deformations can also be reduced to study the deformations of the single-variable function $z^n$ near 0. Adding extra terms of degree less then $n$ results a degree $n$ polynomial, which has $n$ complex roots in general. These complex roots converge to 0 if the coefficients of the extra terms tend to 0. 
 The number of real roots is at most $n$ and at least $1$ if $n$ is odd, and the signs of the roots alternate. 
 As we already noted in the main text, if we allow higher degree terms in the deformation, more roots appear, but these are coming from the infinity, not from zero, in the following sense. Consider $z^n-tz^{n+1}=-t z^n \cdot (z-1/t)$. As $t$ tends to zero, the extra root $1/t$ converges to the infinity. 

Notice that previous paragraph remains true if $z^n$ is replaced by an arbitrary (locally convergent) power series $f(z)$ of order $n$. Any small generic complex deformation of $f$ has $n$ roots coming from $0$ and any small generic real deformation has at most $n$ roots, at least $1$ in the odd case, and their signs alternate. In this sense the local multiplicity can be considered as the generalization of the order of the single-variable power series to higher dimensions.

\subsection*{Example 2}

 The $n$-fold Weyl point, including the double ($n=2$) and triple ($n=3$) Weyl points, appears in many contexts. 
 See, for example, the electronic band structure of multi-Weyl semimetals, as described in \cite{ChenFang_multiweyl}. 
An $n$-fold Weyl point is described by the map
\bean
h(x, y, z) = 
\begin{pmatrix}
\text{Re}(x+iy)^n\\
\text{Im}(x+iy)^n\\
z
\end{pmatrix}.
\eean
Similary to the previous example, we can  omit the identity component $z$ for the calculation to obtain a map $f: (\R^2,0) \to (\R^2,0)$ with the same local degree and local multiplicity as $h$. For this reason, the local degree and the local multiplicity of the 3D double Weyl points are the same as those of the 2D double Weyl point in bilayer graphene (see main text).
Below, we show that the local degree of the $n$-fold Weyl point is $\deg_0 h= \deg_0 f=n$, whereas its local multiplicity is $\mbox{mult}_0 h= \mbox{mult}_0 f= n^2$. 

\emph{Local degree from the constant-deformation method.}
Let us introduce the complex variable $w=x+ iy$.
The equation $ w^n=t$ ($t \neq 0$) has $n$ complex solutions, which implies that $f_t = f - (t,0)$ has $n$ real roots, which in turn implies that $\deg_0 f = \deg_0 h =n$.
Geometrically speaking, this local degree is the winding number of $w^n$ on the complex plane, or, alternatively, the winding number of the map $\frac{h}{|h|}|_{S^{2}}: S^2 \to S^2$, or, alternatively, the winding number of the map $\frac{f}{|f|}|_{S^{1}}: S^1 \to S^1$.

\emph{Local multiplicity from the constant-deformation method.}
Here, we sketch the derivation of the local multiplicity $\mbox{mult}_0 h$ as the local degree $\deg_0 h_\C$ of the complexification. To obtain the local multiplicity, we have to find the complex solutions $x$ and $y$ of the system of equations
\begin{subequations}\label{eq:wnconst}
\bean
    \text{Re}(x+iy)^n &=& t_1, \\
    \text{Im}(x+iy)^n &=& t_2.
\eean
\end{subequations}
These equations can be solved exactly, and the number of complex solutions is $n^2$, implying that the local multiplicity of $h$ is $n^2$. For completeness we will provide the explicit solutions of Eq.~\eqref{eq:wnconst} for $(t_1,t_2)=(t^n,0)$, see Eq.~\eqref{eq:wncomplex1}, but first we introduce the more effective algebraic methods to compute the local multiplicity directly from $h$, without referring to deformations.

\emph{Local multiplicity from the algebraic method.}
We can also find $\mbox{mult}_0 h$ as the dimension $\dim Q_0(h)$ of the local algebra associated to $h$. 
We show this calculation only for $n=2$. 
In this case, it holds that $\text{Re}(x+iy)^n=x^2-y^2$ and $\text{Im}(x+iy)^n=2xy$. Therefore the ideal $I_h \subset \R\{x, y, z\}$ is generated by 
\begin{equation}
    x^2-y^2, \ xy \mbox{ and } z .
\end{equation}

First of all, notice that $x^3$ and $y^3$ are in $I_h$. Indeed, 
\begin{equation}
    x^3=x \cdot (x^2-y^2)+y \cdot xy,
\end{equation}
and similarly for $y^3$.
To identify the residue class 
\begin{equation}
    [f] \in Q_0(h)=\frac{\R\{x, y, z\}}{I_h}
\end{equation} 
of $f \in \R\{x, y, z\}$,
we first reduce $f$ by substituting $xy=z=x^3=y^3=0$, since the residue class of these terms are 0. 
The resulting power series $\widetilde{f}$ has only pure $x$ and $y$ powers up to second degree.  
\begin{equation}
    \widetilde{f}=a_{00}+a_{10}x+ a_{20}x^2+a_{01}y+a_{02}y^2  
\end{equation} 
 and
\begin{equation}
    \widetilde{g}=b_{00}+b_{10}x+ b_{20}x^2+b_{01}y+b_{02}y^2
\end{equation}
are in the same residue class if and only if $\widetilde{f}-\widetilde{g}$ is divisible by $x^2-y^2$, that is, \begin{equation}
    a_{00}=b_{00}, \ a_{10}=b_{10}, \ a_{01}=b_{01} \mbox{ and } a_{20}+a_{02}=b_{20}+b_{02}.
\end{equation}
These 4 parameters can be chosen independently to determine a residue class. In other words, a possible basis of 
\begin{equation}
    Q_0 (h)=\frac{\R\{x, y\}}{I(xy, x^2-y^2)}
\end{equation}  
is formed by the residue classes of $1$, $x$, $y$ and $x^2+y^2$, hence the dimension is 4. 

The same argument can be presented in a more convenient way, by substituting the elements of the ideal by 0. That is,
\begin{equation}
    x^2-y^2=0 \mbox{, } xy=0 \mbox{ and } z=0.
\end{equation}
Then we have the 4 nonzero residue classes
\begin{equation}
    [1],[x],[y],[x^2]=[y^2],
\end{equation}
and we can see that the residue class of every higher order term is zero, e.g.
\begin{equation}
    [x^3]=[x^2] \cdot [x]=[y^2] \cdot [x]=[y \cdot xy]=[y] \cdot [xy]=0.
\end{equation}

The above calculation can be further simplified using the unique factorization in the ring of power series. Define \begin{equation}
    f_1=x, \ f_1=y, \ g_1=x-y, \ g_2=x+y.
\end{equation} 
Then 
\begin{equation}
    h_1=xy=f_1 f_2 \mbox{ and }h_2=x^2-y^2=g_1 g_2
\end{equation} 
are the unique irreducible decompositions of the components of $h$. The multiplicity has the following property, known from the theory of complex plane curve singularities, see e.g. \cite[Chapter 3]{fulton}:
\begin{equation}\label{eq:intmult}
    \dim \left(\frac{\R\{x, y\}}{I(f_1g_1, f_2g_2)}\right)=\sum_{i, j=1}^2 \dim \left(\frac{\R\{x, y\}}{I(f_i, g_i)}\right).
\end{equation} 
In our case, each of the 4 terms are equal to 1, hence the multiplicity is 4.

\begin{figure*}
	\begin{center}
		\includegraphics[width=1.5\columnwidth]{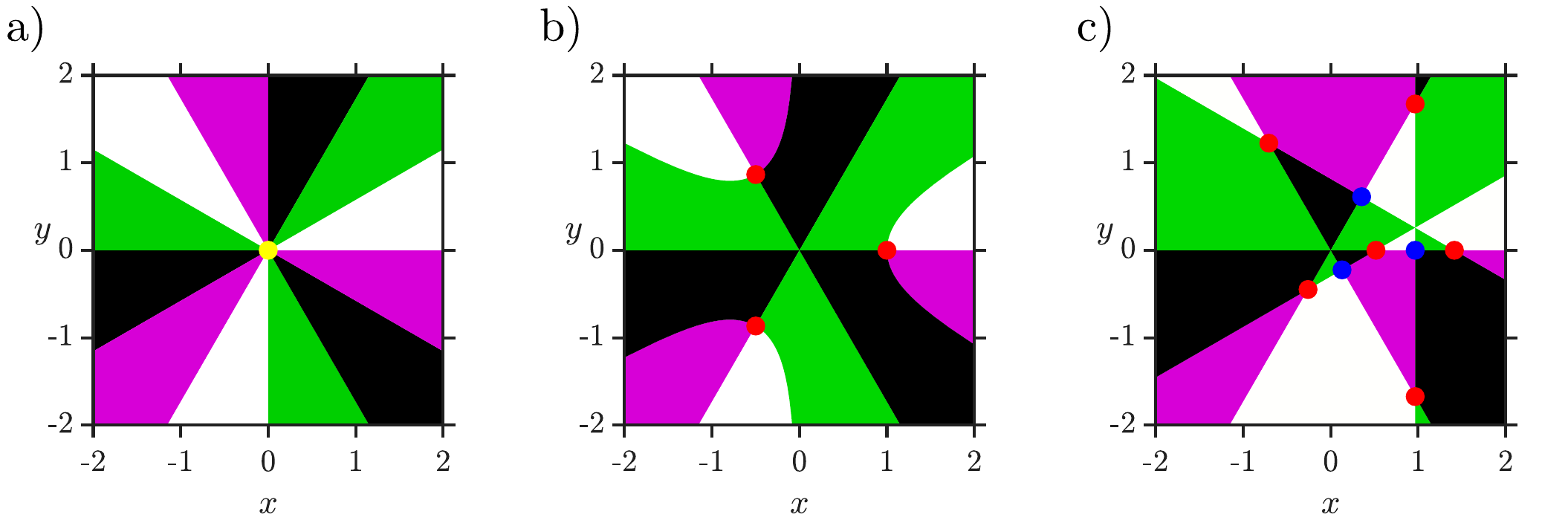}
	\end{center}
	\caption{Birth quota of the 3-fold Weyl point.
	a) 3-fold Weyl point as a non-generic degeneracy point.
	Colored plane shows the sign pattern of the effective Hamiltonian map $h \equiv h_{t=0}$ of Eq.~\ref{eq:wnpert1}.
	For the coloring scheme, see Fig.~1 of the main text.
	b) Three Weyl points (red) born due to deformation of Eq.~\eqref{eq:wnpert1} with $t=1$.
	c) Nine Weyl points (red and blue) born due to deformation of Eq.~\eqref{eq:wnpert2} with $t=1$.
	The multiplicity of the 3-fold Weyl point is 9, i.e., this panel illustrates the case when the number of newborn Weyl points is maximal. 
	\label{fig:w3}}
\end{figure*}

This process also works for greater $n$, e.g for $n=3$
\bean
    \text{Re}(x+iy)^3&=&x^3-3xy^2 \nonumber \\ &=&
    x(x-\sqrt{3}y)(x+\sqrt{3}y),\\
    \text{Im}(x+iy)^3&=&3x^2y-y^3 \nonumber \\ &=&
    y(\sqrt{3}x-y)(\sqrt{3}x+y),
\eean
hence the multiplicity is $3\cdot 3=9$. Compare with Fig~\ref{fig:w3}, where the deformation Eq.~\eqref{eq:wnpert2} pulls apart the the zero loci of the components revealing the 9 generic intersection points.

\emph{Local multiplicity from the `square rule'.} 
The third method uses the relation between the multiplicities of a complex map $g: (\C^m, 0) \to (\C^m, 0) $ and its realification $g_\R: (\R^{2m}, 0) \to (\R^{2m}, 0) $. In this case, \mbox{$\mbox{mult}_0 g_\R=\left(\mbox{mult}_0 g\right)^2$} holds by the proof of \cite[Prop. 2.4.]{EisLev}. Applying this for the map $g(w)=w^n$, we conclude that $\mbox{mult}_0 h=n^2$.

\emph{Example deformations.} Here we provide two special deformations with the explicit location of the Weyl points. The first one is the constant deformation which has the minimal number $|\deg_0 h|$ of real Weyl points, and the second one has the maximal number $\mbox{mult}_0 h$ of real Weyl points, although, in both cases the number of generic complex Weyl points is $\mbox{mult}_0 h$. 

\emph{1st deformation.} Consider the simplest complex deformation of the complex function $w^n=(x+iy)^n$, namely the constant deformation \mbox{$(x+iy)^n-t^n$}, which has $n$ complex roots. Its realification induces the following real deformation of $h$:
\bean\label{eq:wnpert1}
h_t(x,y,z)=\begin{pmatrix}
\text{Re}(x+iy)^n-t^n\\
\text{Im}(x+iy)^n\\
z
\end{pmatrix}.
\eean
Then, the $n$ complex roots of $w^n-t^n=0$ imply $n$ real Weyl points at
\bean\label{eq:wnreal1}
\begin{pmatrix}
x\\
y\\
z
\end{pmatrix}_{\text{WP},k}=t
\begin{pmatrix}
\cos\left(\frac{2\pi k}{n}\right)\\
\sin\left(\frac{2\pi k}{n}\right)\\
0
\end{pmatrix},
\eean
where $0\leq k<n$, with charge $+1$, since the realification of a complex map preserves the orientation.

Furthermore, the complexification of $h_t$ has $n^2$ complex roots at
\bean\label{eq:wncomplex1}
\begin{pmatrix}
x\\
y\\
z
\end{pmatrix}_{\C\text{WP},km}&=&te^{i\pi\frac{2k-m}{n}}
\begin{pmatrix}
\cos\left(\frac{\pi m}{n}\right)\\
\sin\left(\frac{\pi m}{n}\right)\\
0
\end{pmatrix}
\eean
with $0\leq k,m<n$. The real Weyl points in Eq.~\eqref{eq:wnreal1} are the solutions corresponding to the indices satisfying $m \equiv 2k \ (\mbox{mod } n)$, the others are non-real complex Weyl points.

For the $n=3$ case, the non-generic degeneracy point corresponding to $t=0$ is shown in Fig.~\ref{fig:w3}a as the yellow point, and the 3 Weyl points born from the latter, corresponding to $t=1$ are shown in Fig.~\ref{fig:w3}b.

\emph{2nd deformation.} The deformation
\bean\label{eq:wnpert2}
h_t=\begin{pmatrix}
\text{Re}\left(x+iy-te^{i\frac{\pi}{4n}}\right)^n\\
\text{Im}(x+iy)^n\\
z
\end{pmatrix}
\eean
splits the $n$-fold Weyl point into $n^2$ real Weyl points at
\bean\label{eq:wnreal2}
\resizebox{.75\hsize}{!}{$\begin{pmatrix}
x\\
y\\
z
\end{pmatrix}_{\text{WP},km} = t \frac{\sin\left(\frac{\pi}{n}\left(k+\frac{1}{4}\right)\right)}{\sin\left(\frac{\pi}{n}\left(k-n+\frac{1}{2}\right)\right)} 
\begin{pmatrix}
\cos\left(\frac{\pi m}{n}\right)\\
\sin\left(\frac{\pi m}{n}\right)\\
0
\end{pmatrix}$},
\eean
where $0\leq k \leq n-1$ and $1\leq m \leq n$. Their charges are
\bean
Q_{km}=(-1)^{k+m}\cdot\text{sgn}\left(\frac{1}{2}+k-m\right).
\eean

Fig.~\ref{fig:w3}c shows the $n=3$ case for $t=1$. The non-generic degeneracy point splits to 9 real Weyl points.

\subsection*{Example 3}
The first two examples were $(\R^3,0) \to (\R^3,0)$ maps where 1 or 2 components were trivial, therefore, those could be treated as $(\R,0) \to (\R,0)$ and $(\R^2,0) \to (\R^2,0)$ maps effectively. Moreover, the second example suggests the inequality \mbox{$(\deg_0 h)^2 \leq \text{mult}_0 h$}, but this is not true in general. The following example is a $(\R^3,0) \to (\R^3,0)$ map from \cite[Pg. 24]{EisLev}, where all 3 components are non-trivial, and it is also a counterexample for the inequality
\bean\label{eq:mult12}
h(x, y, z) = 
\begin{pmatrix}
xyz\\
x^2-y^2\\
x^2-z^2
\end{pmatrix}.
\eean
This map has local degree $\deg_0 h=4$ and local multiplicity $\mbox{mult}_0 h=12$, as stated in \cite{EisLev}. Note that the inequality $2 |\deg_0 h| \leq \mbox{mult}_0 h$ is proved in \cite{EisLev} if $0$ is a nongeneric root of $h$.

Even though example Eq.~\eqref{eq:mult12} was presented in \cite{EisLev} in a purely mathematical context, a similar map also arises in the electronic band structure context, highlighted in Eq.~(2) of \cite{TiantianZhang}.
Denoting the map defined by their Eq.~(2) as $h_\text{nSOC}$, and assigning $A=B=-1$, it holds that $h_\text{nSOC} = g \circ h$, where $g$ is the linear map
\bean
g(x,y,z) = \frac{1}{2}\begin{pmatrix}
z+y\\
\sqrt{3}(z-y)\\
2x
\end{pmatrix}.
\eean
It is clear that composing a map with a linear transformation $g$ with positive determinant preserves the local degree and the local multiplicity, therefore we have $\deg_0 (g \circ h)=\deg_0 h$ and $\mbox{mult}_0 (g \circ h)=\mbox{mult}_0 h$.

\emph{Local multiplicity from the algebraic method.} The result $\mbox{mult}_0 h = 12$ can be derived easily using the irreducible decomposition of the components of $h$. In fact, \bean
I_h=I(xyz, (x-y)(x+y), (x-z)(x+z)),
\eean
and by the general version of Eq.~\eqref{eq:intmult} we have
\bean
\mbox{mult}_0 h &=& \mbox{mult}_0 (x, x-y, x-z) 
\nonumber \\
&+& \mbox{mult}_0 (x, x-y, x+z)+ \dots \nonumber \\
&+& \mbox{mult}_0 (z, x+y, x+z).
\eean
Here all the 12 terms are 1. For example 
\bean
I(x, x-y, x-z)=I(x, y, z),
\eean
whose multiplicity is 1, since a basis of $\R\{x, y, z\}/I(x, y, z)$ is formed by the residue class of 1.

\emph{Example deformations.} Here we provide two special deformations with the explicit location of the Weyl points. The first one is the constant deformation which has the minimal number $|\deg_0 h|$ of real Weyl points. The second one has the maximal number $\mbox{mult}_0 h$ of real Weyl points, although, in both cases the number of generic complex Weyl points is $\mbox{mult}_0 h$. 

\emph{1st deformation.}
Consider the following constant deformation $h(x, y, z)-(t^3, 0, 0)$, reads
\bean
h_t(x,y,z)=\begin{pmatrix}\label{eq:12min}
xyz-t^3\\
x^2-y^2\\
x^2-z^2
\end{pmatrix}.
\eean
It has 4 real Weyl points at
\bean
(x,y,z)_{\text{WP}}=(\pm t,\pm t,\pm t)\ |\ \text{sgn}(xyz)=1,
\eean
Each has charge $+1$, since the determinant of the Jacobian of $h_t$ is positive everywhere, except the origin. This observation proves that $\deg_0 h=4$. Furthermore, there are 8 non-real complex Weyl points at
\bean
\exp(\pm 2\pi i/3)\cdot (x,y,z)_{\text{WP}}.
\eean
Therefore, the number of complex Weyl points proves that $\mbox{mult}_0 h=12$.

\begin{figure}
	\begin{center}
		\includegraphics[width=1\columnwidth]{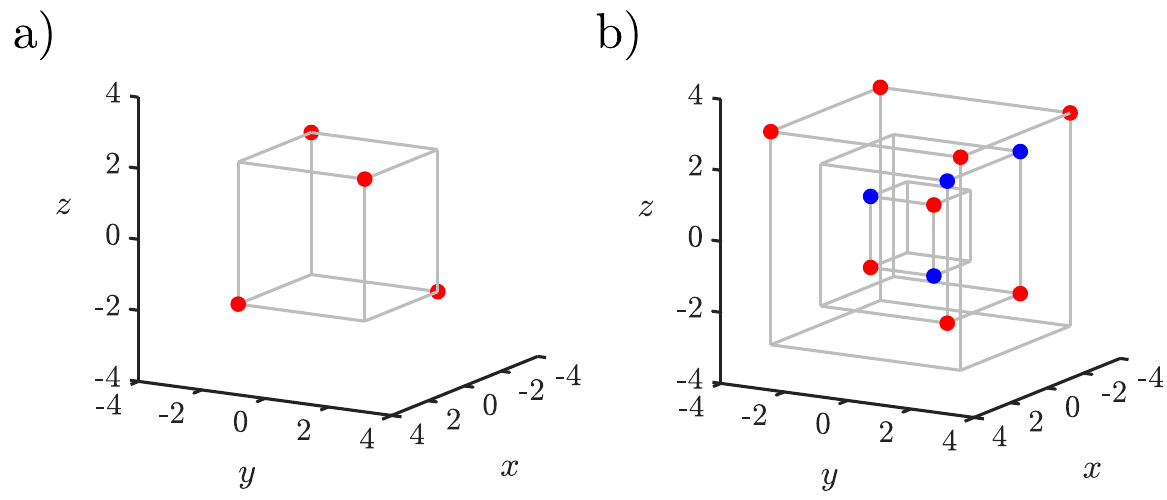}
	\end{center}
	\caption{Birth quota of the map defined in Eq.~\eqref{eq:mult12}.
	a) The minimal number of $\deg_0 h=4$ Weyl points (red) born due to the constant deformation of Eq.~\eqref{eq:12min} with $t=2$.
	b) The maximal number of $\mbox{mult}_0 h=12$ Weyl points (red and blue) born due to the deformation of Eq.~\eqref{eq:12max} with $t=1$. The sum rule for the topological charge is satisfied with 8 Weyl points with charge +1 (red) and 4 Weyl points with charge -1 (blue).
	\label{fig:R3}}
\end{figure}

Fig.~\ref{fig:R3}a shows the 4 real Weyl points at every second corner of a cube with edge length $2t$ around the origin.

\emph{2nd deformation} The deformation
\bean
h_t(x,y,z)=\begin{pmatrix}\label{eq:12max}
(x-t)(y-2t)(z-3t)\\
x^2-y^2\\
x^2-z^2
\end{pmatrix}
\eean
splits the degeneracy into the maximal number of $\mbox{mult}_0 h=12$ real Weyl points with charges $Q$ at 
\bean
(x,y,z)_{\text{WP}}&=&(t,\pm t,\pm t)\hspace{11mm}Q=\text{sgn}(yz),\nonumber\\
(x,y,z)_{\text{WP}}&=&(\pm 2t,2t,\pm 2t)\hspace{6mm}Q=-\text{sgn}(z),\\
(x,y,z)_{\text{WP}}&=&(\pm 3t,\pm 3t,3t)\hspace{6mm}Q=1.\nonumber
\eean
The topological charge sum rule is satisfied with the 8 positively charged Weyl points and the 4 negatively charged Weyl points giving the charge of the unperturbed non-generic degeneracy.

Fig.~\ref{fig:R3}b shows the 12 real Weyl points at corners of 3 concentric cubes, each contributes with 4 corners corresponding to 1 face. 

\subsection*{Example 4}
In the previous three subsections, we computed the local multiplicities of $(\R^3,0) \to (\R^3,0)$ maps describing energy degeneracies of quantum systems. 
In all cases so far, the local multiplicities  were finite. 
In this subsection, we showcase a $(\R^3,0) \to (\R^3,0)$ map that exemplifies that the local multiplicity of an isolated degeneracy can also be infinite. 
This example map is the following:
\bean\label{eq:infty}
h(x, y, z) = 
\begin{pmatrix}
xz\\
yz\\
z^2-\left(x^2+y^2\right)
\end{pmatrix}.
\eean

\emph{Local multiplicity from the algebraic method.} We show that $\mbox{mult}_0 h=\infty$. This follows from the following observations:

a) Since $xz$, $yz$ and $z^2$ are divisible by $z$ (they are 0 if $z=0$), the components of $h$ are contained in the ideal generated by the functions $z$ and $x^2+y^2$. Therefore $I_h \subset I{(z, x^2+y^2)}$. We conclude that 
\bean\label{eq:infty}
 \resizebox{.75\hsize}{!}{$\mbox{mult}_0 h= \dim \frac{\C\{x, y\}}{I_h} \geq \dim \frac{\C\{x, y\}}{I(z, x^2 +y^2)}$}.
\eean

b) We show that the right hand side of Eq.~\eqref{eq:infty} is infinite. 
In particular we show that the residue classes
\bean
[1], [x], [x^2], [x^3], \dots
\in \frac{\C\{x, y\}}{I(z, x^2 +y^2)}
\eean
are linearly independent. Indeed, their finite linear combinations are exactly the residue classes of the one variable polynomials \bean 
a_n x^n + a_{n-1} x^{n-1} + \dots + a_1 x +a_0,
\eean
and these polynomials are not contained in the ideal $I(z, x^2 +y^2)$.

\emph{Complexification.}
According to the infinite multiplicity, $h_\C$ is not finite at 0, i.e. 0 is not isolated in the zero locus $h_\C^{-1}(0)$. Indeed, $x^2+y^2$  in $\C\{x, y\}$ decomposes as 
\bean x^2+y^2=(x+iy)(x-iy).
\eean
Therefore 
the set of the common solutions of $z=0$ and $x^2+y^2=(x+iy)(x-iy)=0$ is the union of the two complex  lines $x+iy=0$ and $x-iy=0$ intersecting at 0.

\emph{Example deformation.}
Since in this case $\mbox{mult}_0 h=\infty$,  there is no birth quota for the number of Weyl points.

\begin{figure}
	\begin{center}
		\includegraphics[width=1\columnwidth]{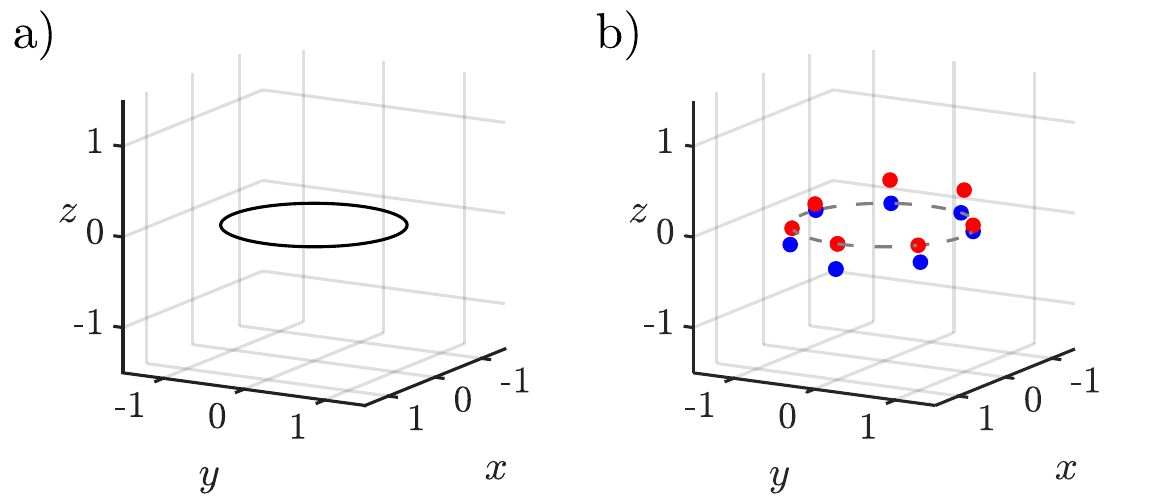}
	\end{center}
	\caption{Splitting a degeneracy point without a birth quota. The map defined in Eq.~\eqref{eq:infty} has multiplicity infinity, thus, it has no birth quota.
	a) The deformation Eq.~\eqref{eq:infty2} with $t=1$ splits the degeneracy point into a nodal circle (black solid line) in the $xy$ plane with radius 1.
	b) The nodal circle is further splitted into Weyl points (red and blue points) with the deformation Eq.~\eqref{eq:infty3} with $n=7$ and $s=1$. This deformation leaves 7 Weyl points on the former circle (chopped gray line) with additional 7 Weyl points nearby giving 14 Weyl points together. It is possible to split the degeneracy directly into $2n$ Weyl points with applying the two deformations together with $s=t$ going from 0 to 1.
	\label{fig:infinity}}
\end{figure}

The deformation
\bean\label{eq:infty2}
h_t(x,y,z)=\begin{pmatrix}
xz\\
yz\\
z^2-\left(x^2+y^2\right)+t^2
\end{pmatrix}
\eean
splits the degeneracy point at the origin into a nodal loop 
\bean
\begin{pmatrix}
x\\
y\\
z
\end{pmatrix}_{\text{loop}}&=&t
\begin{pmatrix}
\cos\varphi\\
\sin\varphi\\
0
\end{pmatrix},\hspace{6mm}0\leq \varphi<2\pi,
\eean
which is \textit{continuously} many degeneracies (see Fig.~\ref{fig:infinity}a). This can be further splitted to \textit{discrete} number of Weyl point without upper bound with the following deformation (compare with Eq.~\eqref{eq:wnpert1})
\bean\label{eq:infty3}
h^n_{t,s}(x,y,z)&=\begin{pmatrix}
xz\\
yz\\
z^2-\left(x^2+y^2\right)+t^2
\end{pmatrix}\nonumber\\
&+s
\begin{pmatrix}
\text{Re}(x+iy)^n-t^n\\
\text{Im}(x+iy)^n\\
0
\end{pmatrix}.
\eean
The second deformation splits the loop everywhere except where the deformation is zero
\bean
\begin{pmatrix}
x\\
y\\
z
\end{pmatrix}_{\text{WP},k}&=&t
\begin{pmatrix}
\cos\left(\frac{\pi k}{n}\right)\\
\sin\left(\frac{\pi k}{n}\right)\\
0
\end{pmatrix},\hspace{6mm}0\leq k<n.
\eean
This leaves $n$ Weyl points at the former loop. Setting $s=t$ and increasing it from 0 to 1 describes a process where the degeneracy point directly splits into at least $n$ Weyl points.

In this case the location of the Weyl points outside the loop is not analytically solvable. By numerical calculations we found that every point has an oppositely charged pair nearby where the condition $h^n_{t,s}(x,y,z)=0$ is satisfied, thus, the total number of Weyl points is $2n$. The topological charge is distributed among the Weyl points in a non-trivial fashion. See Fig.~\ref{fig:infinity}b.

\emph{Local degree.} $\deg_0 h=0$ which can be proven with the deformation
\bean
h_{t}(x,y,z)=\begin{pmatrix}
xz+ty\\
yz-tx\\
z^2-\left(x^2+y^2\right)+t^2
\end{pmatrix}.
\eean
This deformation splits the degeneracy into nothing because $h_{t}(x,y,z)=0$ has no real-valued solution, meaning that the local degree is indeed zero.

\subsection*{Example 5 -- two counterexamples}

Here, we provide two perturbations of the one-variable real function $h(x)=x$, with the number of newborn Weyl points being greater than $\mbox{mult}_0 h=1$. These examples do not invalidate the results of this paper, rather they show that the analytic condition (or the weaker $\mathcal{C}^{\infty}$-condition) \emph{assumed for the deformation} is indeed required for respecting the birth quota. Indeed, in each of these (counter)examples the perturbation $(x, t) \mapsto h_t(x)$, considered as a map of both the configurational and the control parameters, is not analytic, what is more, it is not continuously differentiable. 

\subsubsection{1st counterexample} 
\begin{figure}
	\begin{center}
		\includegraphics[width=1\columnwidth]{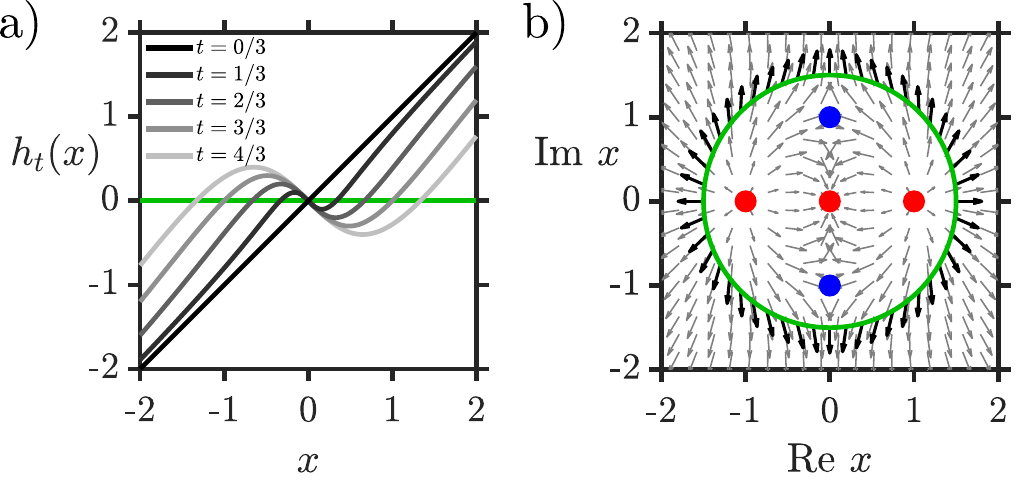}
	\end{center}
	\caption{Perturbation defined in Eq.~\eqref{eq:counter1} of the function $h_0(x)=x$ does not respect the birth quota. a) The function $h_t(x)$ uniformly goes to $x$ as $t$ goes to 0. However, the first derivative is $-1$ for every $t\neq 0$, and $+1$ for $t=0$. In such a process, 3 roots merge in the origin, while keeping the total topological charge (sign) $+1$. b) The complexification $h_{\C t}(x)$ has 2 poles (blue) on the imaginary axis, in addition to the 3 roots (red)  on the real axis, for $t\in\R\setminus \{0\}$. Note that the birth quota is respected in a generalized sense: if we assign negative multiplicity to the poles, then the sum of the multiplicities of the newborn complex roots and poles together equals the original multiplicity.
	\label{fig:counter1}}
\end{figure}

Consider the perturbation shown in Fig.~\ref{fig:counter1}:
\bean\label{eq:counter1}
h_t(x)=x- \frac{2t^2x}{t^2+x^2}=\frac{x\left(x^2-t^2\right)}{x^2+t^2}.
\eean

\emph{Extending to $t=0$ and $x=0$.} For a fixed $t \neq 0$, $h_t(x)$ is an analytic function of $x$ defined in a neighbourhood of $x=0$. For $t=0$, $h_0(x)$ is not defined at $x=0$. However, if $t \to 0$, then $h_t(x)$ converges to $x$ uniformly:
\bean
\lim_{t \to 0} h_t(x) =h(x)=x,
\eean
Hence we can define $h_0(0)=0$ using this limit, consequently $h_0(x)=h(x)=x$ everywhere. Therefore, we could consider $h_t(x)$ as a perturbation of $h(x)$.

\emph{Roots.} $h_t(x)$ has 3 generic roots at
\bean
x_\text{WP}\in\{0,\pm t\},
\eean
therefore the number of the newborn Weyl points from the origin is bigger than the birth quota $\mbox{mult}_0 h=1$. Why does our argument fail in this example? Obviously this perturbation does not satisfy the conditions required for respecting the birth quota. Which condition is not satisfied? 
Is $h_t(x)$ an analytic deformation of $h(x)$? We provide here a detailed analysis of these questions.

\emph{Taylor series of $h_t$ around 0.} 
One can show that the Taylor series of $h_t(x)$ fails to converge to $x$. Observe that 
\bean\label{eq:zoom} 
h_t(x)=t \cdot  h_1\left(\frac{x}{t}\right),
\eean
meaning that varying $t$ scales up the graph of $h_t(x)$. Using this property, the derivatives of $h_t$ at 0 reads
\bean\label{eq:zoom2} 
h_t^{(n)}(0)=t^{1-n} \cdot  h_1^{(n)}\left(0\right),
\eean
for all $t \neq 0$. The first derivative of the perturbed function at 0  is $h'_t(0)=-1$ independently of $t$. This is not equal to the derivative of the unperturbed function $h_0'(0)=1$. Higher derivatives of the perturbed function even diverges as $t$ goes to 0. Therefore, the Taylor series expansion of $h_t(x)$ at $x_0=0$ (with respect to $x$) does not converge to the Taylor series of $h$ at 0. Practically this shows the problem with the perturbation $h_t (x)$  of $h(x)$. 

\emph{Continuity and partial differentials.} For a more fundamental analysis consider the two variable function $\mathcal{H}_1: (x, t) \mapsto h_t(x)$. The problem is that  $\mathcal{H}_1$ is not analytic around zero as a two-variable function. Although it is continuous (if we define $\mathcal{H}_1(0, 0)=0$), it is not continuously differentiable. 

To see the continuity, introduce the polar coordinates $x=r \cos (\alpha)$ and $t=r \sin (\alpha) $, then
\bean
\mathcal{H}_1(r, \alpha)=r \cos (\alpha) \cos (2 \alpha),
\eean
and
\bean
\lim_{(x, t) \to 0} \mathcal{H}_1(x, t)=\lim_{r \to 0} \mathcal{H}_1(r, \alpha)=0.
\eean

To see that $\mathcal{H}_1$ is not continuously differentiable, we show that its partial derivative with respect to $x$ is not continuous at $(0, 0)$. Indeed, we find
\bean
 \partial_x \mathcal{H}_1(x, t)=\frac{x^4+4x^2t^2-t^4}{\left(x^2+t^2\right)^2},
\eean
and writing this in polar coordinates yields
\bean
\partial_x \mathcal{H}_1(r, \alpha)=
\cos (2 \alpha) + \sin^2(2 \alpha).
\eean
This derivative depends on $\alpha$, but it is independent of $r$. Therefore, the limit
\bean
\lim_{(x, t) \to (0, 0)} \left(  \partial_x \mathcal{H}_1(x, t) \right)
=
\lim_{r \to 0} \left(
 \partial_x \mathcal{H}_1(r, \alpha) \right)
\eean
does not exist.

Recall that by definition, $h_t$ is an analytic deformation of a map $h$ if it is the first component $\mathcal{H}_1$ of an analytic unfolding $\mathcal{H}$ with a fixed control parameter value $t$, considered as the function of $x$, see Section~\emph{Birth quota} 
in the main text and Sec.~\ref{app:mult}. The analytic property of $\mathcal{H}$ is used in the proof of Proposition 1. Note that smoothness ($\mathcal{C}^{\infty}$) is also enough to respect the birth quota. Since in this example $\mathcal{H}_1(x, t)$ is not analytic (neither smooth), the perturbation $h_t(x)$ is not an analytic deformation of $h(x)$ according to our definition. 

\emph{Complexification.} It is instructive to investigate the complexification of $h_t(x)$ for $t \neq 0$. It is a meromorphic one-variable function with the 3 generic real roots and two first order
poles at
\bean
x_\text P=\pm it.
\eean
The equation $3-2=1=\mbox{mult}_0 h$  suggest that the poles can be considered as `complex Weyl points with negative sign', and in fact this is the case, as we can see as follows. The function $f(z)=1/z^k$ has a pole of order $k$ at zero. On the unit circle $1/z^k=\overline{z}^k$ (with overline denoting complex conjugation), hence we can change $f$ to $\overline{z}^k$ inside the unit circle to get a continuous function (which can be smoothed along the unit circle). The local degree of this modified function at the origin is $-k$. This result can be interpreted as a pole of order $k$ can be replaced with a root of local degree $-k$. 

Therefore, by this observation, the birth quota \mbox{$\text{mult}_0 h=1$} is respected in a generalized sense. Note that this does not hold in general. Indeed, this example is specific for two reasons: (1) the 1-dimensionality is required for having the isolated poles in the sense used here, and (2) neither the roots nor the poles cross the separator for small values of $t$.

\begin{figure}
	\begin{center}
		\includegraphics[width=1\columnwidth]{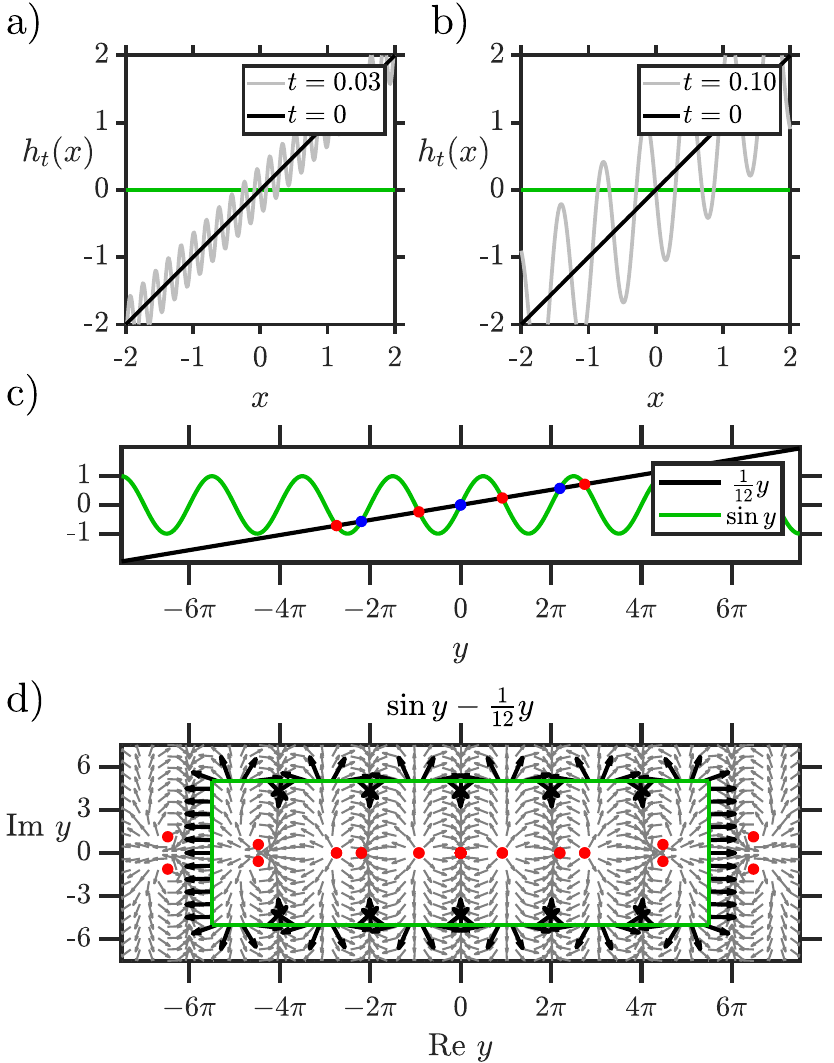}
	\end{center}
	\caption{Perturbation defined in Eq.~\eqref{eq:counter2} breaks the birth quota with an unlimited number of Weyl points depending on the value of $a$. a-b) The perturbed function $h_t(x)$ with $a=1/12$ for $t=0.03$ and $t=0.10$. As $t$ goes to 0, the perturbed function uniformly goes to $x$. c) Graphical solution of the roots of $h_t(x)$. The location of Weyl points is determined by the equation $\sin y=a y$ with $y=x/t$. The number of roots are unlimited as the number of intersections of the sine and a linear function can be arbitrarily large depending on $a$. As $t$ goes to 0, the roots merge at the origin.
	d) The infinitely many complex roots of the function $\sin y-ay$, indicating the infinitely many complex roots of $h_{\C, t}(x)$, with the substitution $y=x/t$. As $t$ goes to 0, infinitely many complex roots cross the separator (green rectangle).
	\label{fig:counter2}}
\end{figure}

\subsubsection{2nd counterexample} A different perturbation of the function $h_0(x)$ can give birth to unlimited number of Weyl points from the origin. For a fixed $0<a$ real number consider
\bean\label{eq:counter2}
h_t(x)=x-\frac{t}{a} \sin \left( \frac{x}{t} \right).
\eean

\emph{Extending to $t=0$.} For $t \neq 0$, $h_t(x)$ is an analytic function of $x$.
Then, $h_0$ can be defined as the uniform limit
\bean
\lim_{t \to 0} h_t (x)=h(x)=x,
\eean
as the sine function is bounded.

\emph{Roots.} $h_t(x)=0$ if and only if
$\sin \left( y \right)=a y$
holds for $y=x/t$. By choosing a sufficiently small $a$ the number of solutions can be arbitrarily large, see Fig.~\ref{fig:counter2}. Take a solution $y_k$, then $x_{\text{WP},k}=ty_k$ is a root of $h_t$. Since $\lim_{t \to \infty} x_{\text{WP},k}=0$, the root $x_{\text{WP},k}$ is born from the origin.

Therefore the formula Eq.~\eqref{eq:counter2} arbitrary many Weyl points born from the origin, while the birth quota is $\mbox{mult}_0 h=1$. The analysis below shows that Eq.~\eqref{eq:counter2} is not a perturbation of $h(x)=x$ according to our definition.

\emph{Taylor series of $h_t$ around 0.}
Similarly to the 1st counterexample, one can show that the Taylor series of $h_t(x)$ does not converge to $x$. Note that the property defined in Eq.~\eqref{eq:zoom} also holds for this example, hence the first derivative of $h_t(x)$ reads
\bean 
h_t'(0)=h_1'(0)=1-\frac{1}{a},
\eean
while $h'(0)=1$. Similarly to the first counterexample the higher derivatives are even diverging.

Clearly this implies that the two-variable map $\mathcal{H}_1: (x, t) \mapsto h_t(x)$ is not analytic (neither smooth). The condition of respecting the birth quota is not satisfied.

\emph{Complexification:} For a fixed $t \neq 0$ the complexification  of the one variable function $h_t(x)$ is holomorphic, and it has infinitely many roots. Each of the roots tends to $0$ in the limit $t \to 0$, in other words, infinitely many roots born from the origin. In contrast with the first counterexample the local multiplicity 1 is not shown by the complexification. It is not surprising, since in this example $h_{\C, t}$ does not converge to $h_{\C}$, as $t \to 0$. In fact, the winding number (degree) of the function restricted to the separator tends to infinity, as $t \to 0$, while the winding of $h_{\C}(x)=x$ is 1.
 
\bibliography{references.bib}
\end{document}